\documentclass[reprint,aps,twocolumn,groupedaddress]{revtex4-1}
\usepackage{color}
\usepackage{graphicx}
\usepackage{enumerate}
\usepackage{amsfonts,amssymb}
\usepackage{amsmath}
\usepackage{braket}
\usepackage{courier}
\usepackage{stmaryrd}
\usepackage{dcolumn}
\usepackage{bbold}
\usepackage{algpseudocode}
\usepackage{algorithm}
\usepackage[all,cmtip]{xy}
\usepackage{hyperref}

\newcommand{\Hilb}{\mathcal{H}}

\newcommand{\vecr}{\mathbf{r}}

\newcommand{\REVISION}[1]{#1}

\begin{document}
\title{A simplified and improved approach to tensor network operators in two dimensions}
\author{Matthew J. O'Rourke}
\affiliation{\footnotesize{Division of Chemistry and Chemical Engineering,
California Institute of Technology, Pasadena, CA 91125, USA}}
\author{Garnet Kin-Lic Chan}
\affiliation{\footnotesize{Division of Chemistry and Chemical Engineering,
California Institute of Technology, Pasadena, CA 91125, USA}}
\bigskip
\date{\today}

\begin{abstract}
Matrix product states (MPS) and matrix product operators (MPOs) are one dimensional
tensor networks that underlie the modern density
matrix renormalization group (DMRG) algorithm. The use of MPOs accounts for the high level
of generality and wide range of applicability of DMRG.
However, current algorithms for two dimensional (2D) tensor network states, known as
projected entangled-pair states (PEPS), rarely employ the associated 2D tensor network
operators, projected entangled-pair operators (PEPOs), due to their computational cost
and conceptual complexity.
To lower these two barriers, we describe how to reformulate a PEPO into a set
of tensor network operators that resemble MPOs by considering the different sets of
local operators that are generated from sequential bipartitions of the 2D system.
The expectation value of a PEPO can then be evaluated on-the-fly using only the action
of MPOs and generalized MPOs at each step of the approximate contraction of the 2D tensor network.
This technique allows for the simpler construction and more efficient energy evaluation of
2D Hamiltonians that contain finite-range interactions, and provides an improved strategy
to encode long-range interactions that is orders of
magnitude more accurate and efficient than existing schemes.
\end{abstract}
\maketitle

\section{Introduction}
\label{sec:intro}
The density matrix renormalization group (DMRG) algorithm~\cite{white1992dmrg,white1993dmrg}
is a popular and successful~\cite{schollwock2005review}
technique for finding the variational ground state
of the Schr\"{o}dinger equation in one spatial dimension (1D).
In its modern form, the variational wave function and the Hamiltonian are represented as
1D tensor networks (TNs), namely matrix product states
(MPS)~\cite{fannes1992,fannes1994,ostlund1995thermodynamic,schollwock2011density}
and matrix product operators
(MPOs)~\cite{verstraete2004matrix,mcculloch2007density,verstraetereview,pirvu2010matrix,
chan2016matrix}. The widespread use of MPOs has allowed for the development of
very general, efficient implementations of the algorithm~\cite{Itensor},
permitting the study of large classes of complex problems in a relatively
black-box manner.

However, the two-dimensional (2D) generalization of MPS, known as projected entangled-pair
states (PEPS)~\cite{nishino1996corner,verstraete2004renormalization,verstraete2006criticality,
orus2014practical}, and their associated ground state
algorithms~\cite{verstraetereview,jordan2008classical,
orus2009simulation,lubasch2014algorithms,corboz2016variational,vanderstraeten2016gradient}
have not yet come close to the same level of generality or range of applicability.
One significant reason for this is that projected entangled-pair operators
(PEPOs)~\cite{crosswhite2008finite,pirvu2010matrix,frowis2010tensor,o2018efficient},
the 2D generalization of MPOs, have been scarcely used in the tensor network
literature to date. The ground state optimization algorithms employed by most authors
instead utilize a significantly less general representation of the Hamiltonian that
is restricted to relatively local interactions~\cite{verstraetereview,jordan2008classical,
orus2009simulation,lubasch2014algorithms,corboz2016variational,
vanderstraeten2016gradient,rezaj1-j2}.
We conjecture that this under-utilization of PEPOs in favor of simpler operator
representations can be
attributed to two facts. Firstly, the construction of a PEPO for an arbitrary 2D Hamiltonian
is more conceptually complicated than the construction of the MPO for the
analogous Hamiltonian in 1D, which itself is still more complicated than building 
the local operators
currently used in 2D simulations. Secondly, when compared to the local operators currently
used in 2D,
the use of PEPOs in a ground state optimization significantly
increases the computational cost of the approximate contraction algorithms for 2D
tensor networks in both the finite~\cite{lubasch_unifying} and infinite
(iPEPS)~\cite{jordan2008classical,orus2009simulation,corboz2014competing} cases.

In this article we describe how to overcome both the
computational and conceptual complexity of using general
tensor network operator representations of the Hamiltonian in 2D algorithms. To do so,
we first briefly summarize the MPO formalism and review some well-known examples that are
central ideas in this work (Section~\ref{sec:MPOs}).
We then introduce a new type of tensor network operator known as a generalized MPO (gMPO),
which is closely related to the traditional MPO (Section~\ref{sec:gMPOs}).
Next we show how to reformulate the calculation of the expectation value of a general PEPO
into a series of operations involving only MPOs and gMPOs, which we call the boundary gMPO
method (Section~\ref{sec:PEPOviagMPO}).
Since the language of MPOs is much better known
than that of PEPOs, this reformulation serves to simplify the construction of general 2D Hamiltonians
for most readers. In Section~\ref{sec:results} we demonstrate this simplicity by
reporting the explicit forms of the gMPOs for various representative types of 2D Hamiltonians.
We also show that the new scheme sacrifices no accuracy compared to the explicit usage of a PEPO,
while providing large speedups in computational time.
In addition, a new scheme for efficiently constructing and evaluating
Hamiltonians with long-range interactions is shown to be many orders of magnitude more
accurate and efficient than existing PEPO-based
approaches~\cite{o2018efficient,li2019generalization,lin2019low}.


\section{Matrix Product Operators (MPOs)}
\label{sec:MPOs}

Since many detailed and comprehensive presentations of MPOs already exist~\cite{crosswhite2008finite,
pirvu2010matrix,frowis2010tensor,schollwock2011density,chan2016matrix}, this section will
simply contain a brief overview in order to establish notation, as well as some simple examples
which we will call upon in later sections.

\subsection{Overview}

Consider a 1D system which has been discretized into $L$ localized sites, each with a local Hilbert
space $\Hilb_i$ of dimension $d_i$. A general operator $\hat{O}$ acting on such a system can be written
as,
\begin{equation}
    \hat{O} = \sum_{ \{ \hat{o}_i \} } O^{o_1 o_2 ... o_L} \hat{o}_1 \hat{o}_2 ... \hat{o}_L,
    \label{eqn:MPO1}
\end{equation}
where $\{\hat{o}_i\}$ is the set of local operators acting on $\Hilb_i$ and $O$ is a rank-$L$ tensor with indices
$o_i$ whose dimensions are equal to the cardinality of their respective set $\{ \hat{o}_i \}$.
$O$ contains the weights associated with all possible configurations of the local operators
$\hat{o}_i$.

By fixing the indices, a specific element $O^{o_1 o_2 ... o_L}$ of the tensor $O$
can then be decomposed into a product of matrices $W[i]$,
\begin{equation}
    O^{o_1 o_2 ... o_L} = \sum_{ \{\alpha \} } W_{\alpha_1}^{o_1}[1] ~
    W_{\alpha_1 \alpha_2}^{o_2}[2] ~ ... ~ W_{\alpha_{L-1}}^{o_L}[L],
    \label{eqn:MPO2}
\end{equation}
where $\alpha$ indexes the so-called ``virtual'' or ``auxiliary'' indices which are introduced to perform the
matrix multiplication. In Eq.~\eqref{eqn:MPO2} the $o_i$ are simply labels, intended to indicate that each matrix
$W[i]$ is chosen specifically so that their product reproduces the element $O^{o_1 o_2 ... o_L}$.
However, if the labels are all reinterpreted as their corresponding indices from Eq.~\eqref{eqn:MPO1}, then
we see that the full tensor $O$ can be reconstructed
as the contraction over rank-3 tensors $W[i]$.

Similar to MPS, this decomposition of a rank-$L$ tensor into $L$ rank-3 tensors is motivated by
the fact that most operators of interest do not contain general $L$-body interactions,
but instead are usually limited to few-body terms. This means that, while in general this decomposition could be exponentially
expensive, often the $O$ tensor is
quite sparse and such a transformation can be a highly efficient way to represent the full tensor.

\begin{figure}[t]
    \centering
    \includegraphics[width=0.45\textwidth]{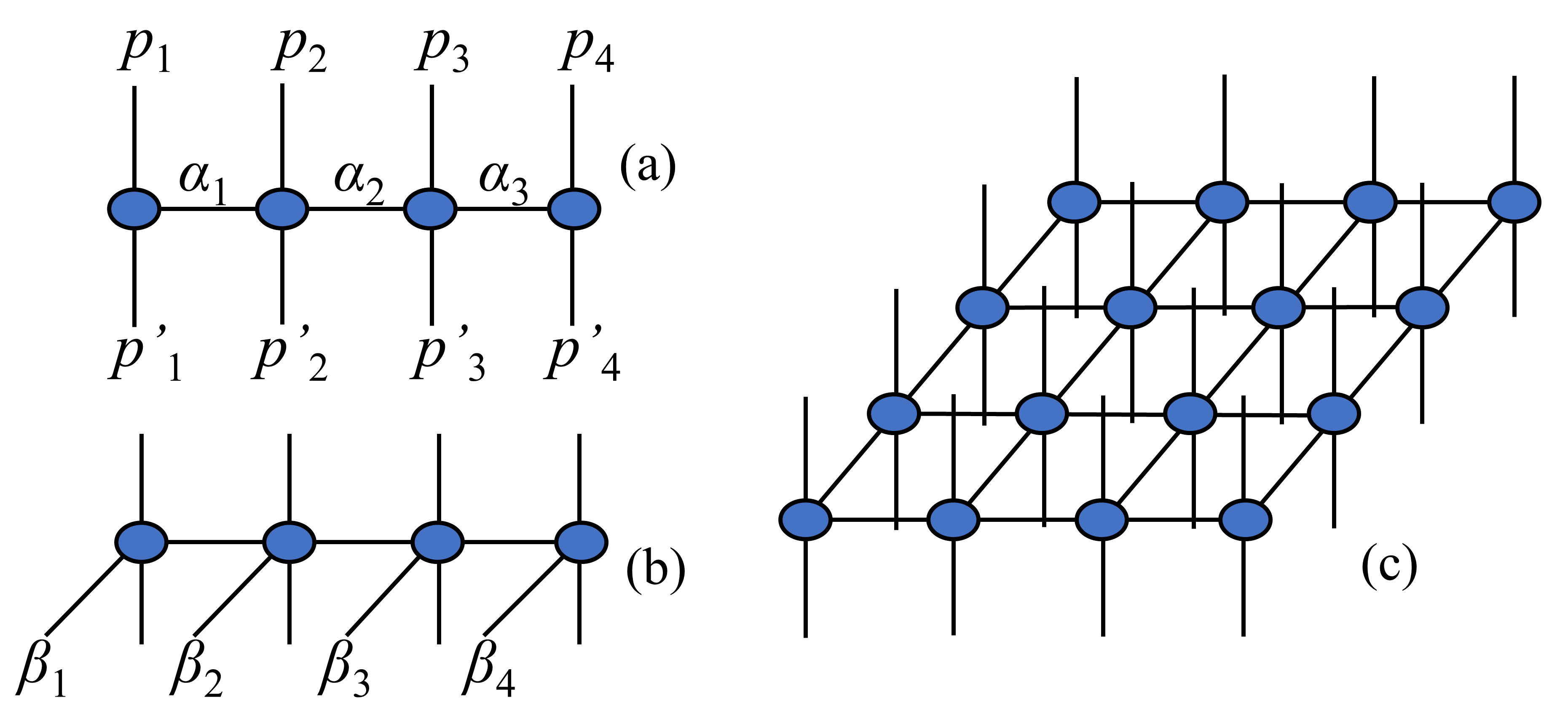}
    \caption{Tensor network diagrams of (a) an MPO, (b) a gMPO, and (c) a PEPO.}
    \label{fig:mpo_gmpo_pepo}
\end{figure}

It is common and frequently useful to associate the operators $\hat{o}_i$ with their corresponding
coefficient tensor $W[i]$ according to,
\begin{equation}
    \hat{W}_{\alpha_{i-1} \alpha_i}[i] = \sum_{o_i} W^{o_i}_{\alpha_{i-1} \alpha_i}[i] ~ \hat{o}_i.
    \label{eqn:MPO3}
\end{equation}
This yields matrices $\hat{W}[i]$ in which every element is a $d_i \times d_i$ local operator acting on
$\Hilb_i$. The full operator $\hat{O}$ is thus reconstructed via simple matrix multiplication,
\begin{equation}
    \hat{O} = \sum_{\{ \alpha \}} \hat{W}_{\alpha_1}[1] ~ \hat{W}_{\alpha_1 \alpha_2}[2]
     ~ ... ~ \hat{W}_{\alpha_{L-1}}[L],
    \label{eqn:MPO4}
\end{equation}
and the set of matrices $\{ \hat{W}[i] \}$ are referred to as the MPO representation of $\hat{O}$.
This form of an MPO is commonly used throughout the literature, and will be heavily utilized in
the remainder of this work.

We will now relate the MPO form in Eq.~\eqref{eqn:MPO4} to the common diagrammatic representation, as
seen in Fig.~\ref{fig:mpo_gmpo_pepo}. Since every element of $\hat{W}_{\alpha_{i-1} \alpha_i}[i]$ is itself a
$d_i \times d_i$ matrix, each individual numerical element can be exposed by introducing two new
indices $p_i$ and $p'_i$, each of dimension $d_i$. By fixing each of $\alpha_{i-1}, \alpha_i, p_i,$ and $p'_i$,
the expression $(\hat{W}_{\alpha_{i-1} \alpha_i}[i])_{p_i p'_i}$ yields a single number. More commonly
written as $\hat{W}_{\alpha_{i-1} \alpha_i}^{p_i p'_i} [i]$, the correspondence to the rank-4 tensors
shown in MPO diagrams becomes apparent. The new indices $p_i$ and $p'_i$ are the so-called ``physical''
indices, which map the action of the local operators onto the corresponding site tensors of an MPS.

\subsection{Examples}
\label{sec:MPOexamples}

Frequently the operator that one wants to encode as an MPO is a Hamiltonian $\hat{H}$, so that the
DMRG algorithm can be used to find its ground state in the form of an MPS. Here we will explicitly
write out the well-known matrices $\hat{W}[i]$ which make up the MPO representations of several common
Hamiltonians consisting of 1- and 2-body terms. There are multiple techniques that can be used
to derive these matrices, each with their own conventions and notation, but in this work we will
remain agnostic to these different languages in an attempt to make the presentation in the following
sections as conceptually simple and widely accessible as possible. To do so, we will simply
refer back to these explicit examples.
In lieu of derivations we will point to helpful references for readers who do not already have a
preferred technique for understanding the form of MPO matrices.

\subsubsection{Nearest-neighbor interactions}
\label{sec:nnMPO}
Consider a system of $L$ sites, which are indexed by $i \in \{ 1,2,...,L\}$,
and a Hamiltonian consisting of local terms and
nearest-neighbor interactions of the form $\hat{H} = \sum_{i=1}^L \hat{C}_i +
\sum_{i=1}^{L-1} \hat{A}_i  \hat{B}_{i+1}$. In the MPO literature this Hamiltonian 
is usually written with $\hat{B} = \hat{A}$ so that the interaction is symmetric 
and $\hat{H}$ is Hermitian, however in this paper we will always keep the operators 
distinct for purposes of notational clarity, even though this means that some Hamiltonians 
under consideration will be non-Hermitian when $\hat{B} \neq \hat{A}$. The MPO matrices 
for this Hamiltonian, denoted $\hat{W}_{NN}$, are given by,
\begin{gather}
    \hat{W}_{NN}[1] = \left( \begin{array}{ccc}
    \hat{C} & \hat{A} & \hat{I}
    \end{array}\right),
    ~ \hat{W}_{NN}[L] = \left( \begin{array}{ccc}
    \hat{I} & \hat{B} & \hat{C}
    \end{array} \right) ^{T}, \nonumber \\
    \hat{W}_{NN}[i] = \left( \begin{array}{ccc}
    \hat{I} & \hat{0} & \hat{0} \\
    \hat{B} & \hat{0} & \hat{0} \\
    \hat{C} & \hat{A} & \hat{I}
    \end{array} \right),
    \label{eqn:NNmpo}
\end{gather}
where $\hat{I}$ is the identity operator and $\hat{0}$ is the zero operator.

If instead the interaction is symmetric so that $\hat{H} = \sum_{i=1}^L \hat{C}_i +
\sum_{i=1}^{L-1} (\hat{A}_i  \hat{B}_{i+1} +\hat{B}_i  \hat{A}_{i+1})$, then the MPO
matrices $(\hat{W}_{NN-sym})$ are given by,
\begin{gather}
    \hat{W}_{NN-sym}[1] = \left( \begin{array}{cccc}
    \hat{C} & \hat{A} & \hat{B} & \hat{I}
    \end{array}\right), \nonumber \\
    \hat{W}_{NN-sym}[L] = \left( \begin{array}{cccc}
    \hat{I} & \hat{B} & \hat{A} & \hat{C}
    \end{array} \right) ^{T}, \nonumber \\
    \hat{W}_{NN-sym}[i] = \left( \begin{array}{cccc}
    \hat{I} & \hat{0} & \hat{0} & \hat{0} \\
    \hat{B} & \hat{0} & \hat{0} & \hat{0} \\
    \hat{A} & \hat{0} & \hat{0} & \hat{0} \\
    \hat{C} & \hat{A} & \hat{B} & \hat{I}
    \end{array} \right).
    \label{eqn:snnMPO}
\end{gather}

In general, for an exact MPO representation of a Hamiltonian $\hat{H}$,
the required bond dimension of the MPO
matrices is $D = 2 + b \cdot r$, where $r$ is the maximum distance over which interactions occur
and $b$ is the number of unique operators that act ``first'' in the interactions.
This is reflected in Eq.~\eqref{eqn:NNmpo} where $r=1$ and $b=1$, and in Eq.~\eqref{eqn:snnMPO}
where $r=1$ and $b=2$.
To understand these patterns, as well as the form of the MPO matrices in this section, we
recommend Ref.~\cite{crosswhite2008finite}.

\subsubsection{Exponentially decaying interactions}
\label{sec:expMPO}
One important exception to the above result is the MPO representation of a Hamiltonian which
has long-range interactions that decay exponentially, such as $\hat{H} = \sum_i \hat{C}_i +
\sum_{i<j} e^{-\lambda (j-i)} \hat{A}_i \hat{B}_j$. Here we have introduced a second index $j$
which runs from $i+1$ to $L$. Despite the fact that $r=L$ in this case, the Hamiltonian has an exact,
compact representation with $D=3$ MPO matrices $(\hat{W}_{exp})$ of the form,
\begin{gather}
    \hat{W}_{exp}[1] = \left( \begin{array}{ccc}
    \hat{C} & \hat{A} & \hat{I}
    \end{array}\right),
    ~ \hat{W}_{exp}[L] = \left( \begin{array}{ccc}
    \hat{I} & e^{-\lambda} \hat{B} & \hat{C}
    \end{array} \right) ^{T}, \nonumber \\
    \hat{W}_{exp}[i] = \left( \begin{array}{ccc}
    \hat{I} & \hat{0} & \hat{0} \\
    e^{-\lambda} \hat{B} & e^{-\lambda} \hat{I} & \hat{0} \\
    \hat{C} & \hat{A} & \hat{I}
    \end{array} \right).
\end{gather}
Refs.~\cite{crosswhite2008applying,pirvu2010matrix,frowis2010tensor,li2019generalization}
provide insight into
why this is possible for the unique case of exponential interactions.

A special case of this representation, which will prove useful in later sections, is when
$\lambda = 0$. The Hamiltonian then has long-range interactions between every pair of sites
but the strength of the interactions are all the same, $\hat{H} =
\sum_i \hat{C}_i + \sum_{i<j} \hat{A}_i \hat{B}_j$. We will denote this special case with
its own MPO notation: $\hat{W}_{uniform}$.

Much like before, if the interactions are symmetric so that $\hat{H} = \sum_i \hat{C}_i +
\sum_{i \neq j} e^{-\lambda |j-i|} \hat{A}_i \hat{B}_j$ (where now both $i,j \in \{1,...,L\}$),
the MPO matrices become,
\begin{gather}
    \hat{W}_{exp-sym}[1] = \left( \begin{array}{cccc}
    \hat{C} & \hat{A} & \hat{B} & \hat{I}
    \end{array}\right), \nonumber \\
    \hat{W}_{exp-sym}[L] = \left( \begin{array}{cccc}
    \hat{I} & e^{-\lambda} \hat{B} & e^{-\lambda} \hat{A} & \hat{C}
    \end{array} \right) ^{T}, \nonumber \\
    \hat{W}_{exp-sym}[i] = \left( \begin{array}{cccc}
    \hat{I} & \hat{0} & \hat{0} & \hat{0} \\
    e^{-\lambda} \hat{B} & e^{-\lambda} \hat{I} & \hat{0} & \hat{0} \\
    e^{-\lambda} \hat{A} & \hat{0} & e^{-\lambda} \hat{I} & \hat{0} \\
    \hat{C} & \hat{A} & \hat{B} & \hat{I}
    \end{array} \right).
\end{gather}
Again, we will give the special case of $\lambda = 0$ its own notation, $\hat{W}_{uniform-sym}$, which
will prove useful in the coming sections.

\subsubsection{General two-body long-range interactions}
\label{sec:LRMPO}
As mentioned previously, exact MPO representations
of Hamiltonians with general long-range interaction
coefficients $\hat{H}_{gen} = \sum_i \hat{C}_i + \sum_{i<j} V_{ij} \hat{A}_i \hat{B}_j$ require a bond dimension
which is proportional to $L$~\cite{frowis2010tensor}. However, if $V_{ij}$ is a smoothly
decaying function of the distance between two sites, $V_{ij} = f(j-i)$, then highly
accurate \textit{approximate} MPO representations of $\hat{H}_{gen}$ can often be found which have
finite, constant bond dimensions. The traditional technique is to fit $f(j-i)$ by a sum of
exponentials~\cite{crosswhite2008applying,pirvu2010matrix},
\begin{equation}
    f(j-i) = \sum_{k=1}^K a_k e^{-\lambda_k (j-i)}.
    \label{eqn:expFit}
\end{equation}
This yields an MPO representation of $\hat{H}_{gen}$ with bond dimension $K+2$, where the MPO
matrices take the form,
\begin{gather}
    \hat{W}_{K-exp}[1] = \left( \begin{array}{cccccc}
    \hat{C} & a_1 \hat{A} & a_2 \hat{A} & \cdots & a_K \hat{A} & \hat{I}
    \end{array}\right), \nonumber \\
    \hat{W}_{K-exp}[L] = \left( \begin{array}{cccccc}
    \hat{I} & e^{-\lambda_1} \hat{B} & e^{-\lambda_2} \hat{B} & \cdots & e^{-\lambda_K }\hat{B} & \hat{C}
    \end{array} \right) ^{T}, \nonumber \\
    \hat{W}_{K-exp}[i] = \left( \begin{array}{cccccc}
    \hat{I} & \hat{0} & \hat{0} & \cdots & \hat{0} & \hat{0} \\
    e^{-\lambda_1} \hat{B} & e^{-\lambda_1} \hat{I} & \hat{0} & \cdots & \hat{0} & \hat{0} \\
    e^{-\lambda_2} \hat{B} & \hat{0} & e^{-\lambda_2} \hat{I}  & \cdots & \hat{0} & \hat{0} \\
    \vdots & \vdots & \vdots & \ddots & \hat{0} & \hat{0} \\
    e^{-\lambda_K} \hat{B} & \hat{0} & \hat{0} & \cdots & e^{-\lambda_K} \hat{I} & \hat{0} \\
    \hat{C} & a_1 \hat{A} & a_2 \hat{A} & \cdots & a_K \hat{A} & \hat{I}
    \end{array} \right).
    \label{eqn:multiExpMPO}
\end{gather}
The accuracy of the representation $\{ \hat{W}_{K-exp} \}$ is determined by the quality of the fit
in Eq.~\eqref{eqn:expFit}.

Although this is often a reasonably accurate approach,
several more sophisticated techniques have been developed
in recent years which are based on the singular
value decomposition (SVD) of blocks of $V_{ij}$~\cite{chan2016matrix,slicedbasis}.
These methods also work most effectively when $V_{ij}$ is a
smooth function of the distance, but they are able
to fit more general functions $f$ that may be challenging to represent directly with exponentials
like those in Eq.~\eqref{eqn:expFit}~\cite{chan2016matrix}.
They also can be a bit more efficient, producing a higher
accuracy representation of $\hat{H}_{gen}$ with a smaller bond dimension
than Eq.~\eqref{eqn:multiExpMPO}~\cite{slicedbasis}.

\begin{figure}[t]
    \centering
    \begin{tabular}{c}
    \includegraphics[width=0.40\textwidth]{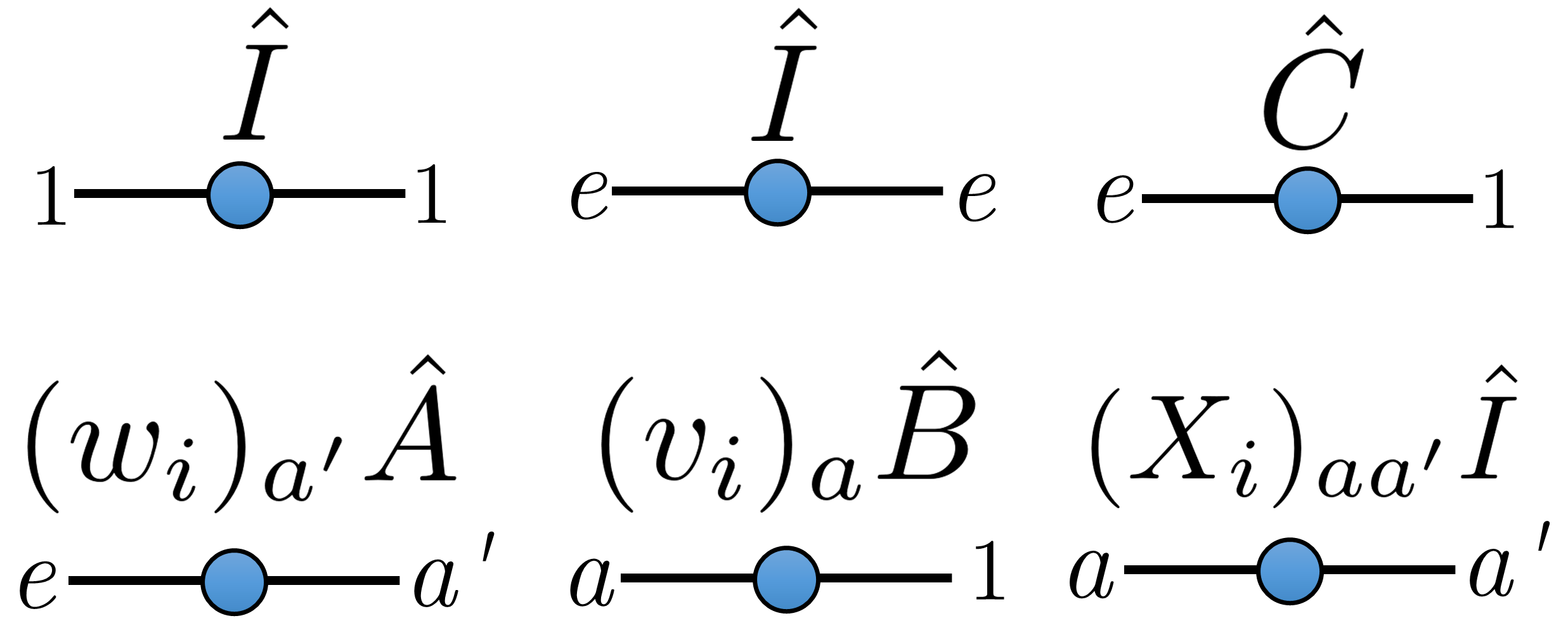} \\
    (a) \\
    \vspace{0.2cm} \\
    \includegraphics[width=0.40\textwidth]{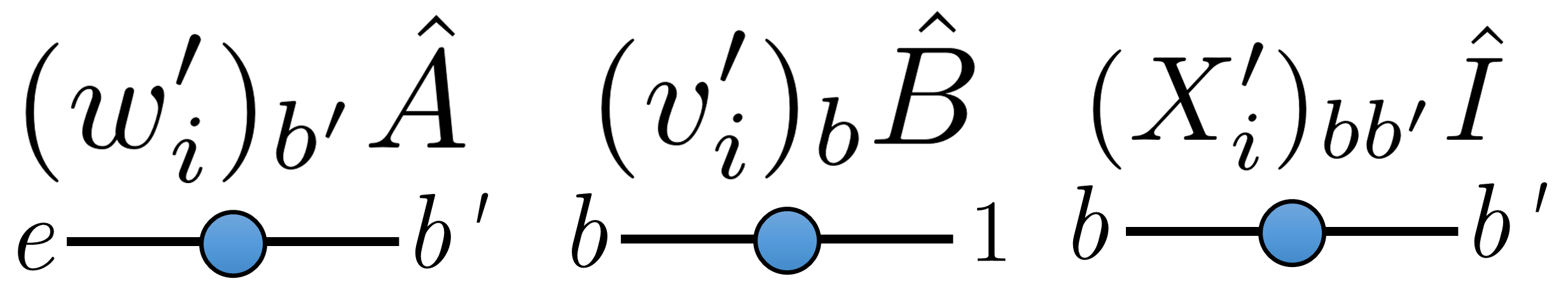}\\
    (b)
    \end{tabular}
    \caption{A set of tensor network diagrams that represent the elements of the operator-valued MPO
    matrix $\hat{W}_{gen}[i]$ (a), along with the additional elements needed for
    $\hat{W}_{gen-sym}[i]$ (b). Here we assume $\hat{W}_{gen}[i]$
    is a $(2+l_i) \times (2+ r_i)$ matrix and $\hat{W}_{gen-sym}[i]$ is a $(2+2 l_i) \times
    (2 + 2 r_i)$ matrix. We use the symbol ``1'' to denote the first value of a given index, $e$ to
    denote the final value of a given index, $a$ to denote the
    set of values ranging from 2 to $l_i +1$, $b$ to denote the set of values ranging from $l_i+2$
    to $2l_i+1$, $a'$ to denote the set of values ranging from 2 to $r_i+1$, and $b'$ to denote
    the set of values ranging from $r_i + 2$ to $2r_i +1$. This index labelling corresponds directly
    to the expressions in Equations~\eqref{eqn:generalMPO} and \eqref{eqn:generalMPOsymm}.
    }
    \label{fig:mpo-gen}
\end{figure}

In this work, we utilize the technique described in Ref.~\cite{slicedbasis}. The basic idea
is that the MPO matrices $\hat{W}_{gen}$ for the general Hamiltonian $\hat{H}_{gen}$ can be written as,
\begin{equation}
    \hat{W}_{gen}[i] = \left( \begin{array}{ccc}
    \hat{I} & \hat{0} & \hat{0} \\
    (v_i)_{a} \hat{B} & (X_i)_{a a'} \hat{I} & \hat{0} \\
    \hat{C} & (w_i)_{a'} \hat{A} & \hat{I}
    \end{array} \right),
    \label{eqn:generalMPO}
\end{equation}
where $\vec{v}_i$ is a column vector of coefficients that has length $l_i$ and is indexed by $a$,
$X_i$ is an $l_i \times r_i$ matrix of coefficients indexed by $a$ and $a'$, and $\vec{w}_i$ is a
row vector of coefficients that has length $r_i$ and is indexed by $a'$, yielding a
$(2+l_i) \times (2+r_i)$ MPO matrix. 
\REVISION{We write the indexed elements of $\vec{v}_i$, $\vec{w}_i$, and $X_i$ in 
Eq.~\eqref{eqn:generalMPO} to remind the reader of the shape of these quantities.}
For clarity, tensor network diagrams for this
matrix are given in Fig.~\ref{fig:mpo-gen}(a). If the coefficients contained in $\hat{W}_{gen}[i]$
can be, to a good approximation, related to the coefficients contained in
$\hat{W}_{gen}[i+1]$ by a linear transformation, then the MPO matrices for each site
can be successively generated by finding the correct linear transformation on the
coefficients contained in the MPO matrix on the previous site. These linear transformations
can be found by taking SVDs of certain blocks of the upper triangle of $V_{ij}$. It is observed
in~\cite{slicedbasis} that if $V_{ij}$ is a smooth function of the distance $|j-i|$,
the transformations are often compact (i.e. their dimensions do not scale with $L$)
and highly accurate because sub-blocks of the upper triangle of $V_{ij}$ are low-rank.
These ideas are developed in full detail in the supplementary
information of Ref.~\cite{slicedbasis}~\footnote{It should be noted that in the referenced
article there is a typo in the explicit expressions for the compressed MPO matrices. The
local operators associated with the $X$ block of each matrix should be $\hat{I}$, not $\hat{n}$,
as in Eq.~\eqref{eqn:generalMPO}.}.

The form of this MPO matrix can be viewed as a direct generalization of $\hat{W}_{uniform}$. The
``coefficients'' in adjacent $\hat{W}_{uniform}$ matrices can be related to each other via the simplest possible
linear transformation ($X_{1\times 1}=1$, $\vec{v} = \vec{w} = 1$) because all the
interactions are of identical strength and thus all sub-blocks of $V_{ij}$ are rank 1.
However, when the interaction coefficients vary with distance and the sub-blocks of the
upper triangle of $V_{ij}$ are rank-$l$, the single $\hat{I}$ in the center of $\hat{W}_{uniform}$
gets generalized to an $l \times l$ block $X_{l\times l} \hat{I}$ in $\hat{W}_{gen}$. By extension,
$\vec{v}$ and $\vec{w}$ undergo the same generalization. The MPOs $\hat{W}_{exp}$ and
$\hat{W}_{K-exp}$ are special, simple cases of this generalization.

As a final note, if the interactions in the general Hamiltonian $\hat{H}_{gen}$ become
symmetric so that $\hat{H} = \sum_i \hat{C}_i + \sum_{i \neq j} V_{ij} \hat{A}_i \hat{B}_j =
\sum_i \hat{C}_i + \sum_{i<j} V_{ij} \hat{A}_i \hat{B}_j + \sum_{j<i} V_{ij} \hat{B}_j \hat{A}_i$,
then the general MPO matrices become,
\begin{equation}
    \hat{W}_{gen-sym}[i] = \left( \begin{array}{cccc}
    \hat{I} & \hat{0} & \hat{0} & \hat{0} \\
    (v_i)_{a} \hat{B} & (X_i)_{a a'} \hat{I} & \hat{0} &\hat{0} \\
    (v'_i)_{b} \hat{A} & \hat{0} & (X'_i)_{b b'} \hat{I} & \hat{0} \\
    \hat{C} & (w_i)_{a'} \hat{A} & (w'_i)_{b'} \hat{B} & \hat{I}
    \end{array} \right).
    \label{eqn:generalMPOsymm}
\end{equation}
Tensor network diagrams representing this matrix are given in Fig.~\ref{fig:mpo-gen}.
Here we have introduced the additional indices $b$ and $b'$ to index the new vectors
$\vec{v'}_i$, $\vec{w'}_i$ and the new matrix $X'_i$, as described in Fig.~\ref{fig:mpo-gen}.
If we have the additional property that interaction coefficients themselves are symmetric,
$V_{ij} = V_{ji}$, then the above expression can be simplified according to:
$\vec{v'}_i = \vec{v}_i$, $\vec{w'}_i=\vec{w}_i$, $X'_i = X_i$.


\section{PEPO expectation value via generalized MPOs}

\subsection{Generalized MPOs (gMPOs)}
\label{sec:gMPOs}
In order to relate the contraction of PEPOs to the well-known 1D MPOs described in
Section~\ref{sec:MPOexamples}, we must first introduce the notion of a generalized MPO (gMPO).
In a gMPO, the operator-valued MPO matrices $\hat{W}[i]$ are elevated to rank-3 tensors,
which will be indicated by the addition of a virtual index $\beta_i \in \{1,2,...,g\}$.
The new operator-valued, rank-3
gMPO tensors will be denoted by $\hat{M}_{\beta_i}[i]$. Exposing all the indices explicitly, this
gives a rank-5 tensor $M_{\alpha_{i-1} \alpha_i \beta_i}^{p_i p'_i} [i]$, which is shown in
diagrammatic form in Fig.~\ref{fig:mpo_gmpo_pepo}.

The basic notion of a gMPO is that for each value of $\beta_i$, a different MPO \textit{matrix}
$\hat{W}[i]$ can be encoded in the gMPO tensor.
In the simplest case $\beta_i$ only takes a single value ($g=1$)
and thus every gMPO tensor can only represent a single MPO matrix, reducing the gMPO back to
a regular MPO. If instead $\beta_i$ takes two values ($g=2$), then every tensor can represent two
different MPO matrices, and the gMPO can encode $2^L$ different 1D MPOs. In practice, however,
the $\beta_i$ are not ``free'' indices but are instead summed over in the final network just
like the $\alpha$ indices in a regular MPO (see Eq.~\eqref{eqn:MPO4}). The proper notion
of a gMPO is therefore as a tensor network operator that can represent a sum of many regular 1D
MPOs after the $\beta_i$ are appropriately summed over. This formulation is useful because it
provides a flexible framework in which operators in regular MPOs can be
coupled with other operators that act
``outside'' of the 1D domain of the regular MPO. In general it allows for the complete coupling of
two distinct MPOs into one, however in this work we only utilize a simpler special case
in which specific local operators are coupled together. Much like how a local operator
on site $i$ can be coupled to a local operator on site $i+1$ by summing over the index
$\alpha_i$ in a regular MPO, we use the gMPO formalism to couple a local operator that acts
``below'' site $i$
to the local operators on site $i$ by performing an appropriate sum over $\beta_i$.

\begin{figure}[t]
    \centering
    \includegraphics[width=0.49\textwidth]{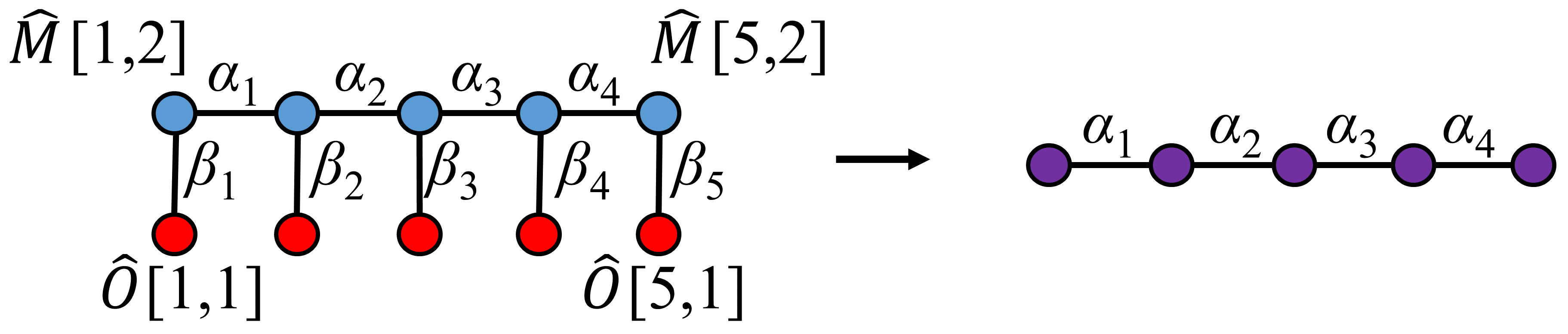}
    \caption{A gMPO-based representation of the two-row example Hamiltonian in
    Section~\ref{sec:gMPOs} for a $2\times 5$ system. \textit{Left}: The gMPO tensors (blue)
    appear on the sites in row 2, while the complementary operator vectors (red) appear on
    the sites in row 1. Physical indices are suppressed for simplicity. \textit{Right}: The
    resulting tensor network along row 2 (an MPO)
    after contractions over the $\beta$ indices have been performed.}
    \label{fig:gmpo_example}
\end{figure}

For clarity, let us consider a simple example. Given a 2D system of size $2 \times L$
consisting of two rows with $L$ sites each, we can label each site by $(i,y)$, where
$i \in \{1, ..., L \}$ as usual and $y \in \{1,2\}$, as depicted in Fig.~\ref{fig:gmpo_example}.
Consider the Hamiltonian
$\hat{H} = \hat{H}_1 + \hat{H}_2 = \sum_i ( \hat{A}_{i,1} \hat{B}_{i,2} + \hat{A}_{i,2} \hat{B}_{i+1,2} )$,
where there are nearest-neighbor interactions between row 1 and row 2 ($\hat{H}_1$), as well as
nearest-neighbor interactions within row 2 ($\hat{H}_2$). This Hamiltonian can be represented by a
simple gMPO ($\hat{M}$) acting on row 2 along with the complementary operators ($\hat{O}$) that
act locally on the sites in row 1.

Since there are no interactions between sites in row 1, the
operators $\{ \hat{O}[i,1] \}$ that are applied in this row take the form of vectors,
like those at the ends of a regular MPO,
but applied along the $\beta_i$ index instead of $\alpha$ (see Fig.~\ref{fig:gmpo_example}),
\begin{equation}
    O_{\beta_i}^{p_{i,1} p'_{i,1}}[i,1] \to \hat{O}_{\beta_i}[i,1] =
    \left( \begin{array}{cc}
    \hat{I}_{i,1} & \hat{A}_{i,1}
    \end{array}\right).
\end{equation}
To couple these operators with the local operators in row 2, as well as to encode
the nearest-neighbor interactions within row 2, gMPO tensors can be used in row 2. They take the form,
\begin{gather}
    \hat{M}_1[i,2] = \hat{W}_{NN}[i], \nonumber \\
    \hat{M}_2[1,2] = \left( \begin{array}{ccc}
    \hat{B}_{1,2} & \hat{0} & \hat{0}
    \end{array}\right),
    ~ \hat{M}_2[L,2] = \left( \begin{array}{ccc}
    \hat{0} & \hat{0} & \hat{B}_{L,2}
    \end{array} \right) ^{T}, \nonumber \\
    \hat{M}_2[1<i<L,2] = \left( \begin{array}{ccc}
    \hat{0} & \hat{0} & \hat{0} \\
    \hat{0} & \hat{0} & \hat{0} \\
    \hat{B}_{i,2} & \hat{0} & \hat{0}
    \end{array} \right),
    \label{eqn:gMPOtensors_ex}
\end{gather}
where $\hat{W}_{NN}$ is from Section~\ref{sec:nnMPO} (with $\hat{C} = \hat{0}$). The reason why
the matrix $\hat{M}_2[i,2]$ takes this form can be understood by explicitly considering
what happens during the contraction over $\beta_i$ for a given column $i$.
\begin{gather}
    \sum_{\beta_i} \hat{O}_{\beta_i}[i,1] ~ \hat{M}_{\beta_i}[i,2]  = \nonumber \\
    \hat{I}_{i,1} \cdot
    \left(\begin{array}{ccc}
     \hat{I}_{i,2} & \hat{0} & \hat{0} \\
    \hat{B}_{i,2} & \hat{0} & \hat{0} \\
    \hat{0} & \hat{A}_{i,2} & \hat{I}_{i,2}
    \end{array}\right)
    +
    \hat{A}_{i,1} \cdot
    \left(\begin{array}{ccc}
     \hat{0} & \hat{0} & \hat{0} \\
    \hat{0} & \hat{0} & \hat{0} \\
    \hat{B}_{i,2} & \hat{0} & \hat{0}
    \end{array}\right) \nonumber \\
    =
    \left(\begin{array}{ccc}
    \hat{I} & \hat{0} & \hat{0} \\
    \hat{B}_{i,2} & \hat{0} & \hat{0} \\
    \hat{A}_{i,1} \hat{B}_{i,2} & \hat{A}_{i,2} & \hat{I}
    \end{array}\right).
    \label{eqn:gMPOex}
\end{gather}

The resulting tensor network operator now looks like a regular MPO along row 2 (see Fig.~\ref{fig:gmpo_example}),
and the form of its matrices looks very similar to $\hat{W}_{NN}$ (Eq.~\eqref{eqn:NNmpo}), which
encodes non-symmetric nearest neighbor interactions. The only difference is that in the place
of $\hat{C}$, the 1-body on-site term in Section~\ref{sec:nnMPO}, there is now the inter-row
interaction term $\hat{H}_1$ for column $i$. Thus, if these MPO matrices are now all contracted together
along the $\alpha$ indices in row 2, we will exactly recover all the terms in our original two row
Hamiltonian.

The function of $\hat{M}_2[i,2]$ is thus evident: it couples the inter-row interactions
into an intra-row MPO matrix in a consistent manner with the structure of the intra-row MPO. Without
$\hat{M}_2$, the action of $\hat{A}_{i,1}$ could not be selectively coupled into specific matrix elements
of $\hat{M}_1$. Thus, the form of $\hat{M}_2$ can be simply determined based on an understanding of the structure of
the ``in-row'' MPO matrix $\hat{M}_1$; namely, to which matrix elements the ``external'' operators should couple.
Although this formalism may appear unnecessarily general
in the context of this simple example, its full utility will become apparent in the subsequent
sections as more complicated Hamiltonians are considered.

\subsection{Evaluation of PEPO expectation values using gMPOs}
\label{sec:PEPOviagMPO}

\begin{figure}[t]
    \centering
    \includegraphics[width=0.46\textwidth]{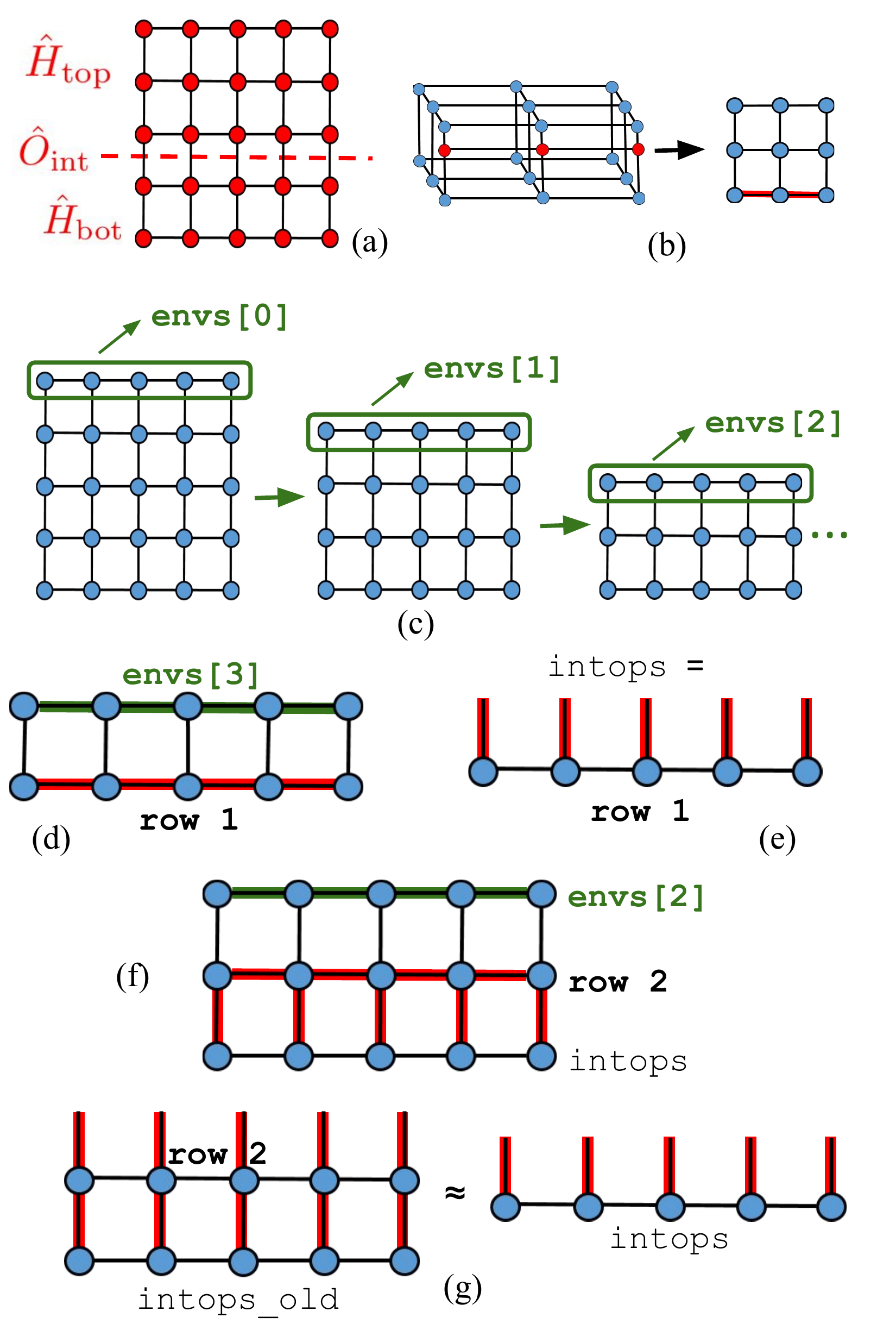}
    \caption{The first full iteration of the boundary gMPO algorithm for a $5 \times 5$ PEPS.
    (a) A $ 5 \times 5$ PEPO (with physical indices that are suppressed)
    that is bipartitioned by a cut between rows 2 and 3. (b) A useful diagrammatic definition:
    when a flat square lattice TN diagram is drawn with some red bonds and some
    black bonds, the red corresponds to where a tensor network operator has been sandwiched between
    the bra and ket. Black bonds contain just bra and ket virtual indices. Figures (c)-(g) are the
    diagrams that directly correspond to the algorithm steps~\ref{step:precompute}-\ref{step:intops},
    respectively (see Section~\ref{sec:PEPOviagMPO}). Green bonds are used to denote the
    pre-computed environments from step~\ref{step:precompute}. }
    \label{fig:p2gmpo}
\end{figure}

To this point, the Hamiltonians under consideration have acted on lattices that are
either strictly or quasi- one dimensional.
In this section we will present an algorithm that utilizes the gMPO formalism
to evaluate the expectation value of fully 2D Hamiltonians with
the same level of generality as PEPOs, but with simpler and more familiar concepts.
This presentation will focus on the case of a finite $L_x \times L_y$ rectangular lattice,
but prospects for its extension to the infinite case will be discussed in
Section~\ref{sec:conclude}. The concepts for this technique begin with consideration
of the three subsets of local operators that
are distinguished by a bipartitioning of the system. Namely, given the full system Hamiltonian
$\hat{H}$ represented by a localized structure such as a PEPO and
a horizontal bipartition of it (as depicted in Fig.~\ref{fig:p2gmpo}(a)), all the
local operators in $\hat{H}$ can be grouped into three mutually exclusive groups:
(i) those for which there are interactions between
sites that are all below the line ($\hat{H}_{\mathrm{bot}}$), (ii) all above
the line ($\hat{H}_{\mathrm{top}}$), or (iii) those for which interactions occur across the
line ($\hat{O}_{\mathrm{int}}$). This decomposition,
\begin{equation}
\hat{H} = \hat{H}_{\mathrm{bot}} + \sum_{ij} h_{ij} \hat{O}_i \hat{O}_j +
\hat{H}_{\mathrm{top}},
\end{equation}
where $i$ indexes sites below the partition, $j$ indexes sites above the partition, and
$h_{ij}$ contains the coefficients for the interactions that get ``cut'', is a familiar
concept in 1D for the analysis of MPOs and is the basis of an efficient implementation
of the DMRG algorithm~\cite{chan2016matrix}. In 2D, it allows for the evaluation of
$\langle \psi | \hat{H} | \psi \rangle$ on-the-fly using gMPOs.

To see how, first consider the contraction of the finite, 2-layer, 2D tensor network corresponding to
 $\langle \psi | \psi \rangle$ for some PEPS $\ket{\psi}$ using the ``boundary MPS''
method~\cite{lubasch_unifying}. Starting from the bottom, 
the first point of reference is \texttt{row 1} and as the contraction progresses, it shifts upward
to \texttt{row 2}, then \texttt{row 3}, etc. During this process the Hamiltonian can be
successively partitioned along
with the reference row of the norm contraction,
so that the first line lies between \texttt{row 1} and \texttt{row 2}, then the next is between
\texttt{row 2} and \texttt{row 3}, etc. Using this idea, the total energy
$\langle \psi | \hat{H} | \psi \rangle$ can be accumulated as follows
(shown graphically in Fig.~\ref{fig:p2gmpo}):

\begin{enumerate}

\item \label{step:precompute} Pre-compute all the partial contractions of $\langle \psi | \psi \rangle$
using the boundary method, starting from the top with \texttt{row $L_y$} and working downward.
They should be stored as $\{$\texttt{envs[0]}, ..., \texttt{envs[$L_y-2$]}$\}$
(Fig.~\ref{fig:p2gmpo}(c)).

\item \label{step:botMPO}
Construct an MPO which contains all the 1-body terms in $\hat{H}$ that act locally in \texttt{row 1}
as well as all the interactions between sites in \texttt{row 1}. In other words, this should be
the MPO
representation of $\hat{H}_{\mathrm{bot}}$ when the partition is between \texttt{row 1} and
\texttt{row 2}. Apply this MPO between the bra and
ket tensors of \texttt{row 1}, and evaluate
$E_{\mathrm{bot}} = \langle \psi | \hat{H}_{\mathrm{bot}} | \psi \rangle$ by contracting this partial TN
with \texttt{envs[$L_y-2$]} (Fig.~\ref{fig:p2gmpo}(d)).

\item \label{step:first_intops} Construct complementary operator vectors which contain the local operators
$\hat{O}_{\mathrm{int}}$ that act in \texttt{row 1} but have interactions with sites above
\texttt{row 1} (as in Section~\ref{sec:gMPOs}).
Apply these vectors between the corresponding
\texttt{row 1} ket and bra tensors along the vertical bonds.
This partial TN will be called \texttt{intops} (Fig.~\ref{fig:p2gmpo}(e)).

\item \label{step:gMPO} Shift the partition line up by 1 row (in general, now in between
\texttt{rows} $y$ and $y+1$).
Construct a gMPO to be applied in \texttt{row $y$}
that encodes all the terms in the new $\hat{H}_{\mathrm{bot}}$
that have not already been evaluated.
Apply the gMPO between the \texttt{row $y$} bra and ket tensors, and contract this
TN with \texttt{intops} (below)
and \texttt{envs[$L_y-y-1$]} (above). Add the resulting scalar to $E_{\mathrm{bot}}$ to obtain a new
$E_{\mathrm{bot}}$, which now accounts for all the terms in
$\langle \psi | \hat{H}_{\mathrm{bot}} | \psi \rangle$ given the new partition position.
For clarity, the case immediately following
step~\ref{step:first_intops} would be when $y=2$.
To accumulate the proper terms, this gMPO should include
interactions within \texttt{row 2}, as well as
all the interactions between sites in \texttt{row 2} and sites in the rows beneath it,
which is just \texttt{row 1} for now ($y=2$ case shown in Fig.~\ref{fig:p2gmpo}(f)).

\item \label{step:intops} Construct an updated (approximate) \texttt{intops}.
This step can be understood as iteratively building up MPOs along the vertical bonds.
First a complementary operator \textit{matrix} (which is just an MPO matrix) is constructed
for each column, which relates the $\hat{O}_{\mathrm{int}}$
in a given column of \texttt{row} $y-1$ to the $\hat{O}_{\mathrm{int}}$ in 
the same column of \texttt{row} $y$. This is exactly like how a
regular MPO matrix relates the operators on site $x-1$ to the operators on site $x$.
Then these complementary operator matrices are applied between each of the bra and
ket tensors of \texttt{row} $y$ along the vertical indices. This row can then be contracted
with the old \texttt{intops} and its horizontal bond dimension can be compressed according
to the boundary method contraction routine. This yields a new approximate
\texttt{intops} that contains the action of all the local operators $\hat{O}_{\mathrm{int}}$ that
lie below the partition when it is between \texttt{rows} $y$ and $y+1$ ($y=2$ case shown in
Fig.~\ref{fig:p2gmpo}(g))

\item Iterate steps~\ref{step:gMPO} and~\ref{step:intops} until the top
of the PEPS is reached.
When the final gMPO is applied
to \texttt{row $L_y$} and contracted with \texttt{intops},
the expectation values of all the terms in $\hat{H}$ will have been tallied in the
running total $E_{\mathrm{bot}}$.
\end{enumerate}

Given a Hamiltonian with general interactions of the form $\hat{A}_i \hat{B}_{j}$, where $i<j$,
the big picture of this algorithm (which we will call the ``boundary gMPO'' method
for future reference) can
be succinctly summarized as follows:
To compute $\langle \psi | \hat{H} | \psi \rangle$, we think
about classifying terms in $\hat{H}$ into 3 non-mutually exclusive groups according to the
bipartition of a PEPO between \texttt{rows} $y$ and $y+1$. Group (1) contains terms where
$\hat{A}$ and $\hat{B}$ are both below the partition. Group (2) contains terms where $\hat{A}$
is below the partition but $\hat{B}$ is somewhere above it. Group (3) contains terms
where $\hat{A}$ and $\hat{B}$ are both below the previous partition (when it was between
\texttt{rows} $y-1$ and $y$). At each iteration of the algorithm, we first compute
$\langle \psi | \hat{H} | \psi \rangle$ for the set of terms in the difference (1) - (3)
by contracting a
gMPO with \texttt{intops}, and then we construct a new \texttt{intops} for the next iteration
that accounts for all the terms in (2) by slightly modifying the previous \texttt{intops}.

This can be viewed as a ``decomposed'' contraction of the
expectation value of a PEPO. As the partition is iteratively shifted upwards,
MPOs are sequentially constructed and applied tensor-by-tensor along
the vertical bonds and gMPOs are applied along the horizontal bonds
in order to ``extract'' the expectation values of the terms in $\hat{H}_{\mathrm{bot}}$, as it is
defined based on the current progress of the contraction. When explicitly contracting
the expectation value of a PEPO, the boundary tensors accumulate the identical terms but they
are not fully evaluated until the entire contraction is complete. By extracting the ``completed''
terms along the way, the boundary gMPO method allows for the energy evaluation of the same set of
general 2D Hamiltonians that can be represented by PEPOs while only invoking MPOs and gMPOs. Since
the ideas for constructing MPOs, and thus also gMPOs, are more familiar and well-established
in the literature than PEPOs, we expect that this will be a useful conceptual simplification.

Additionally,
this formulation leads to a reduction in computational cost because \texttt{intops} can
always be constructed with operator virtual indices pointing only in the vertical direction~\footnote{
Although this is always possible, it is not required. It may be the case that for some Hamiltonians
not explicitly considered in this work, allowing horizontal operator virtual indices in \texttt{intops}
results in a more efficient representation}. When
compared to the contraction of a PEPO, the cost of boundary absorption and compression
(the time-dominant step; step~\ref{step:intops} and Fig.~\ref{fig:p2gmpo}(g) above) is reduced
because the boundary tensors no longer contain any operator virtual indices along the horizontal bonds.
This decreases the cost of boundary absorption by a factor of $D_{\mathrm{op}}^4$ and compression
by a factor of $D_{\mathrm{op}}^6$ (where $D_{\mathrm{op}}$ is the virtual bond dimension of the
PEPO/vertical bond dimension of \texttt{intops} operators)~\footnote{These factors are determined under the assumption that
the bond dimension $\chi$ of the boundary MPS during boundary method contraction~\cite{lubasch_unifying}
must be proportional to $D_{\mathrm{op}}$ for accurate results when using a full PEPO~\cite{o2018efficient}}.

In the context of a variational~\cite{corboz2016variational,vanderstraeten2016gradient}
ground state optimization of a PEPS with respect to the Hamiltonian $\hat{H}$,
this algorithm fits very nicely within the framework of the newly-developed differentiable programming
techniques for tensor networks~\cite{xiang2019differentiable}.
Since the expectation values of different sets of operators
are evaluated during different iterations, each iteration of
steps~\ref{step:gMPO} and \ref{step:intops} can be differentiated separately. This allows for the
gradient of the energy to also be computed on-the-fly as the energy itself is being computed,
leading to a highly efficient computational formulation.


\section{Results}
\label{sec:results}

\begin{table}[t]
    \centering
    \begin{ruledtabular}
    \begin{tabular}{c|cccc}
         & $\hat{H}_{NN}$ & $\hat{H}_D$ & $\hat{H}_{LRNC}$ & $\hat{H}_{LRAC}$    \\
        \hline
	$D=2$, $\chi=5$  & 2.52 & 3.70  &     3.66        &        18.6         \\
        $D=2$, $\chi=10$ & 6.94 & 11.9  &     11.6        &        18.4         \\
        $D=2$, $\chi=20$ & 13.2 & 27.9  &     27.2        &        19.1         \\
        $D=3$, $\chi=15$ & 20.5 & 39.2  &     37.5        &        1.63         \\
        $D=3$, $\chi=30$ & 25.3 & 52.6  &     51.7        &        1.36         \\
        $D=3$, $\chi=40$ & 24.0 & 50.9  &     51.2        &        1.41         \\
        $D=4$, $\chi=15$ & 19.0 & 33.5  &     34.2        &                     \\
        $D=4$, $\chi=30$ & 27.8 & 59.3  &     60.3        &                     \\
        $D=4$, $\chi=40$ & 32.7 & 62.8  &     62.5        &                     \\
    \end{tabular}
    \end{ruledtabular}
    \caption{
    The average computational speedups of the boundary gMPO algorithm over PEPO-based expectation
    value calculations for a representative set of 2D Hamiltonians (Eqns.~\eqref{eqn:ham-2d-nn},
    \eqref{eqn:ham-2d-diag}, \eqref{eqn:ham-2d-lrnc}, and \eqref{eqn:ham-2d-lrac}).
    The gMPO-based scheme is generically and significantly faster than the PEPOs for all the
    Hamiltonians except the one with long-range interactions mediated by a distance-dependent
    potential (LRAC). The reported numbers are averages taken
    over multiple calculations for each of multiple different trial wavefunctions:
    PEPS ground states for the $8 \times 8$
    AFM Heisenberg model and FM transverse field Ising model ($h=3.5$).
    $D$ denotes the bond dimension of the trial PEPSs and $\chi$ denotes the maximum boundary
    bond dimension used during contraction~\cite{lubasch_unifying}.
    Both algorithms were implemented in a straightforward manner
    in order to compare their runtimes as fairly as possible.
    This data should be used in conjunction with Fig.~\ref{fig:no_coeff_acc}.
    }
    \label{tab:PEPOspeedups}
\end{table}

In this section, we will present the explicit constructions of the MPOs and gMPOs
needed to implement the boundary gMPO algorithm described in Section~\ref{sec:PEPOviagMPO} for various
types of 2D Hamiltonians. From the set of Hamiltonians that we explicitly describe, we expect
that the construction of most other Hamiltonians of potential interest will be conceptually
straightforward. We will also demonstrate the speed and accuracy of the new algorithm,
and compare it to the performance of expectation value computations using explicit PEPOs as well as
``brute force'' application of all the Hamiltonian terms separately (this technique is analogous to
the current technique used in 2D simulations, as mentioned in Section~\ref{sec:intro}).

\REVISION{In our brute force implementations we do not utilize any caching strategies for
contraction intermediates that are recyclable between the evalutation of multiple different 
Hamiltonian terms. This would lead to a faster routine, and might allow for a more direct
comparison to the boundary gMPO algorithm since it inherently utilizes a (quite limited)
caching strategy. However, while the implementation of the \texttt{envs} intermediates in the
boundary gMPO method is very straightforward, proper caching for the brute force technique is more
complicated, especially for Hamiltonians which include long-range interactions.
To keep the results for all Hamiltonians comparable, we thus always refrain from caching in
the brute force method.}

In all cases we will consider a finite two-dimensional system on a rectangular
lattice of $L_x \times L_y$ sites
labelled $(x,y)$, where $x$ indexes the sites in a row and $y$ indexes the sites in a
column. By the conventions of the previous sections, $(1,1)$ corresponds to the bottom
left corner and $(L_x,L_y)$ to the top right corner. When a sum is taken over all the sites in
the lattice using a single index, such as $\sum_{i=1}^{L_x \times L_y}$, the order in which the
sites are indexed is such that site $(x+1,y)$ always has a larger label number than
site $(x,y)$, and site $(x,y+1)$ also has a larger label number than $(x,y)$. This convention will
be important when restrictions are placed on the sums, such as the condition $i<j$.

\begin{table}[t]
    \centering
    \begin{ruledtabular}
    \begin{tabular}{c|cccc}
         & $\hat{H}_{NN}$ & $\hat{H}_D$ & $\hat{H}_{LRNC}$ & $\hat{H}_{LRAC}$    \\
        \hline
	$D=2$, $\chi=5$  & 60.8 & 118.4 &      1066        &        34.2         \\
        $D=2$, $\chi=10$ & 59.9 & 107.6 &     975.7        &        33.0         \\
        $D=2$, $\chi=20$ & 49.4 & 84.6  &     782.2        &        24.2         \\
        $D=3$, $\chi=15$ & 43.1 & 72.8  &     672.3        &        16.9         \\
        $D=3$, $\chi=30$ & 36.6 & 64.6  &     609.6        &        16.1         \\
        $D=3$, $\chi=40$ & 37.3 & 65.4  &     628.9        &        16.8         \\
        $D=4$, $\chi=15$ & 39.9 & 61.8  &     592.6        &                     \\
        $D=4$, $\chi=30$ & 35.3 & 63.4  &     569.8        &                     \\
        $D=4$, $\chi=40$ & 37.5 & 68.7  &     623.8        &                     \\
        $16 \times 16$, $D=2$, $\chi=20$ &  &  &           &        296.1
    \end{tabular}
    \end{ruledtabular}
    \caption{
    The average computational speedups of the boundary gMPO algorithm over ``brute force'' expectation
    value calculations for a representative set of 2D Hamiltonians (Eqns.~\eqref{eqn:ham-2d-nn},
    \eqref{eqn:ham-2d-diag}, \eqref{eqn:ham-2d-lrnc}, and \eqref{eqn:ham-2d-lrac}).
    The gMPO-based scheme is significantly faster for all the Hamiltonians under consideration,
    especially those which contain long-range interactions (LRNC and LRAC).
    In the brute force
    technique, the Hamiltonian is evaluated term-by-term by explicitly applying each pair of local
    operators.
    Both algorithms were implemented in a straightforward manner
    in order to compare their runtimes as fairly as possible.
    The reported numbers are averages taken in an identical manner to Table~\ref{tab:PEPOspeedups},
    and the parameters $D$ and $\chi$ are also indentically defined.
    This data should be used in conjunction with Fig.~\ref{fig:no_coeff_acc}.
    }
    \label{tab:bruteSpeedups}
\end{table}

\subsection{Local Hamiltonians}

\subsubsection{Nearest-neighbor interactions}
Consider a Hamiltonian with local 1-body terms and non-symmetric nearest-neighbor interactions
of the form,
\begin{equation}
\hat{H}_{NN} = \sum_{i=1}^{L_x \times L_y} \hat{C}_i + \sum_{\langle ij \rangle, i<j}
\hat{A}_{i} \hat{B}_j,
\label{eqn:ham-2d-nn}
\end{equation}
where both $i$ and $j$ index through all $L_x \times L_y$ sites.
The MPO in step~\ref{step:botMPO} of the boundary gMPO algorithm is given by
$\hat{W}_{NN}$ from Eq.~\eqref{eqn:NNmpo}, Section~\ref{sec:MPOexamples}.
The vertical
MPOs that are applied tensor-by-tensor as the algorithm progresses in order to produce \texttt{intops}
are given by,
\begin{gather}
    \hat{O}_{\beta_1}[x,1] =
    \left( \begin{array}{cc}
    \hat{I} & \hat{A}
    \end{array}\right), \nonumber \\
    \hat{O}_{\beta_{y-1} \beta_y}[x, L_y > y > 1] = \left( \begin{array}{cc}
    \hat{I} & \hat{A} \\
    \hat{0} & \hat{0}
    \end{array}\right).
    \label{eqn:intopsTensors1}
\end{gather}
Note that here we use the index label $\beta_y$ to denote its position ($y$) along the vertical bonds
within column. This is a slight abuse of notation when compared to Section~\ref{sec:gMPOs}, 
where the subscript on $\beta$ was used to denote its position ($x$) within in a single row.
A fully consistent notation would require an $x$ and $y$ subscript on every $\beta$, but for all
Hamiltonians under consideration in Section~\ref{sec:results} the vertical MPO matrices 
will be the same for every $x$, so we always suppress the $x$ label 
(and sometimes also the $y$ label when the context is unambiguous) on $\beta$ for simplicity.

Also note that these MPO matrices (Eq.~\eqref{eqn:intopsTensors1}) only need to be of dimension
$2 \times 2$ because each time a new \texttt{intops} is
created, the expectation values of interaction terms with the row above are immediately extracted by
contracting it with an appropriate gMPO. Unlike a typical MPO, we therefore never need to ``complete''
an interaction with a $\hat{B}$ operator in these matrices because that is taken care of in the gMPO.
This eliminates the need for the third row and column to account for $\hat{B}$. In the current case
of nearest neighbor interactions, the bottom row of $\hat{O}_{\beta_{y-1} \beta_y}$ is all $\hat{0}$s because
the action of $\hat{A}$ in \texttt{row} $y$ does not need to be stored once the point of reference is
shifted up to \texttt{row} $y+1$.

The gMPO tensors, used in step~\ref{step:gMPO} to extract the
expectation values of terms in $\hat{H}_{\mathrm{bot}}$,
were given as the example in Eq.~\eqref{eqn:gMPOtensors_ex}. To make the notation
consistent with a fully 2D Hamiltonian, the coordinates of the tensors in that expression should be
transformed according to: $\hat{M}_1[i,2] \to \hat{M}_1[x,y>1]$; $\hat{M}_2[1,2] \to \hat{M}_2[1,y>1]$;
$\hat{M}_2[L,2] \to \hat{M}_{2}[L_x,y>1]$; $\hat{M}_2[i,2] \to \hat{M}_2[L_x > x > 1, y>1]$.
Additionally, for generality
we do not have $\hat{C}=\hat{0}$ in our current example.
In essence, the evaluation of this Hamiltonian's expectation value amounts to performing the same
calculation as the one outlined in the example of Section~\ref{sec:gMPOs}
for every row in the system.

The accuracy of the boundary gMPO algorithm using these tensors to evaluate the expectation
value of the given Hamiltonian
(with $\hat{C} =  \hat{0}, \hat{A} = \hat{B} = \sigma_z$) with respect to various trial PEPS
is shown in Fig.~\ref{fig:no_coeff_acc}. It is almost identically accurate to the brute force
scheme and its accuracy is also very similar to the PEPO-based
implementation in most cases, with the outliers showing an improved accuracy for the gMPOs.
Despite the similar accuracies, using the gMPOs allows for a computational
speedup of up to $\sim 30\times$
over the PEPOs and $\sim 40\times$ over the brute force implementation, as seen in
Tables~\ref{tab:PEPOspeedups}-\ref{tab:bruteSpeedups}.

\subsubsection{Diagonal-neighbor interactions}
\label{sec:diag}
Now consider a Hamiltonian that has local 1-body terms as well as both nearest-neighbor and
diagonal-neighbor interactions, with
strengths $J_1$ and $J_2$ respectively,
\begin{equation}
    \hat{H}_D = \sum_{i=1}^{L_x \times L_y} \hat{C}_i + J_1 \sum_{\langle ij \rangle, i<j}
    \hat{A}_{i} \hat{B}_j + J_2 \sum_{\langle \langle ij \rangle \rangle, i<j} \hat{A}_i \hat{B}_j.
    \label{eqn:ham-2d-diag}
\end{equation}
Although we are again considering the non-symmetric Hamiltonian
construction (denoted by $i<j$) for simplicity, if the interaction operators are
chosen to be symmetric (i.e. $\hat{B} = \hat{A}$) then the given Hamiltonian differs from
the truly symmetric one $\hat{H}_{\mathrm{sym}}$ (i.e. $i<j \to i \neq j$) by a factor of 2 in the interaction
coefficients, $\hat{H}_{\mathrm{sym}}(J_1,J_2) = \hat{H}_D(2J_1, 2J_2)$. If the representation of
$\hat{H}_{\mathrm{sym}}$ is needed when $\hat{B} \neq \hat{A}$, 
it can be determined by using the results in this section
and following the examples in Section~\ref{sec:MPOexamples}.

\begin{figure}[ht!]
    \centering
    \includegraphics[width=0.48\textwidth]{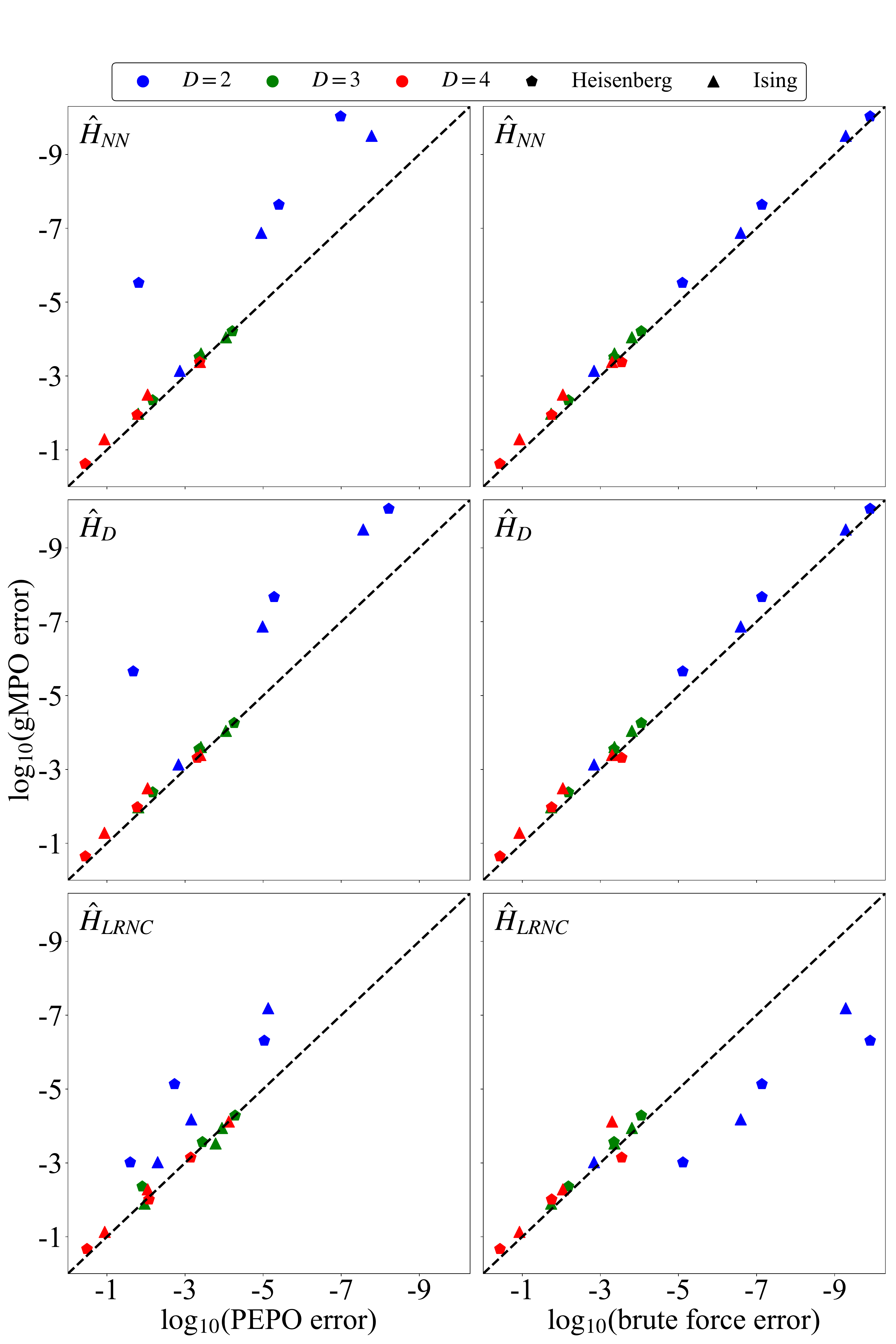}
    \caption{
    The relative error of the gMPO-based expectation values compared to the relative errors
    obtained using both PEPO-based calculations and the ``brute force'' technique of evaluating each
    term in the Hamiltonian separately.
    For the majority of cases tested, all three techniques
    exhibit the same level of accuracy.
    None of these Hamiltonians (Eqns.~\eqref{eqn:ham-2d-nn},
    \eqref{eqn:ham-2d-diag}, \eqref{eqn:ham-2d-lrnc}) contain long-range distance-dependent
    potentials.
    The expectation values
    are calculated with respect to various $8 \times 8$ trial PEPS of bond dimensions $D=2,3,4$.
    A single point compares the relative error
    of gMPOs with either PEPOs or brute force, with each technique using the same
    trial state and $\chi$ value (the boundary bond
    dimension during contraction~\cite{lubasch_unifying}). All errors are
    measured with respect to a brute force evaluation that is highly converged in $\chi$.
    The displayed points are for selected
    values of $\chi$ less than the converged value, in order to compare the levels of accuracy
    that can be obtained with a given computational effort.
    For a full picture of the computational effort,
    this data should be used in conjunction with the speedups reported in
    Tables~\ref{tab:PEPOspeedups}-\ref{tab:bruteSpeedups}, which also include the $\chi$ values
    for each of the points here.}
    \label{fig:no_coeff_acc}
\end{figure}

The MPO in step~\ref{step:botMPO} is again given by $\hat{W}_{NN}$ (Eq.~\eqref{eqn:NNmpo},
Section~\ref{sec:MPOexamples}), and the vertical MPO matrices are still given by
Eq.~\eqref{eqn:intopsTensors1}. The gMPO tensors are,
\begin{gather}
    \hat{M}_1[1,y>1] = \left( \begin{array}{cccc}
    \hat{C} & J_1 \hat{A} & \hat{B} & \hat{I}
    \end{array}\right), \nonumber \\
    \hat{M}_1[L_x,y>1] = \left( \begin{array}{cccc}
    \hat{I} & \hat{B} & \hat{0} & \hat{C}
    \end{array} \right) ^{T}, \nonumber \\
    \hat{M}_1[L_x > x > 1, y>1] = \left( \begin{array}{cccc}
    \hat{I} & \hat{0} & \hat{0} & \hat{0} \\
    \hat{B} & \hat{0} & \hat{0} & \hat{0} \\
    \hat{0} & \hat{0} & \hat{0} & \hat{0} \\
    \hat{C} & J_1 \hat{A} & \hat{B} & \hat{I}
    \end{array} \right), \nonumber \\
    \hat{M}_2[1,y>1] = \left( \begin{array}{cccc}
    J_1 \hat{B} & J_2 \hat{I} & \hat{0} & \hat{0}
    \end{array}\right), \nonumber \\
    \hat{M}_2[L_x,y>1] = \left( \begin{array}{cccc}
    \hat{0} & \hat{0} & J_2 \hat{I} & J_1 \hat{B}
    \end{array} \right) ^{T}, \nonumber \\
    \hat{M}_2[L_x > x > 1, y>1] = \left( \begin{array}{cccc}
    \hat{0} & \hat{0} & \hat{0} & \hat{0} \\
    \hat{0} & \hat{0} & \hat{0} & \hat{0} \\
    J_2 \hat{I} & \hat{0} & \hat{0} & \hat{0} \\
    J_1 \hat{B} & J_2 \hat{I} & \hat{0} & \hat{0}
    \end{array} \right).
    \label{eqn:gMPOdiag}
\end{gather}
These matrices can be understood by noting the similarity between $\hat{M}_1$ and
$\hat{W}_{NN-sym}$ from Eq.~\eqref{eqn:snnMPO}. The only difference is that in $\hat{M}_1$
the entry for $\hat{A}$
in the first column is made to be $\hat{0}$ (and interaction coefficients are included).
This is done to prevent symmetric nearest-neighbor
interactions of the form $\hat{B}_{x-1,y} \hat{A}_{x,y}$ from being included along the gMPO row
(in these coordinates the gMPO is being applied to row $y$).
However, since the sites are ordered in such a way that the (non-symmetric) diagonal-neighbor
interactions occur between site $(x,y-1)$ and sites $(x-1,y)$, $(x+1,y)$, we still want to
include the action of $\hat{B}$ ``on the left'' on site $(x-1, y)$. This is exactly what the
form of $\hat{W}_{NN-sym}$ is designed to do.

If the $\hat{M}_1$ matrices were the only ones included in the gMPO, then this action of $\hat{B}_{x-1,y}$
``on the left'' would never be utilized due to the $\hat{0}$ in place of $\hat{A}_{x,y}$ in the first column.
However, $\hat{M}_2$ couples the action of $\hat{A}_{x,y-1}$ (from \texttt{intops})
into the two typical locations of $\hat{A}$ in
$\hat{W}_{NN-sym}$ (and also multiplies by $J_2$).
This allows the ``on the left'' action of $\hat{B}_{x-1,y}$ to interact
with the action of $J_2 \hat{A}_{x,y-1}$, which is exactly the diagonal interaction that we want to include.
$\hat{M}_2$ also couples $J_2 \hat{A}_{x,y-1}$ into the same position as $J_1 \hat{A}_{x,y}$ in $\hat{M}_1$, which
allows for the nearest-neighbor horizontal interaction and diagonal-neighbor interaction ``to the right''
to be accounted for simultaneously. Specifically, after the $\beta_{y-1}$ indices have been appropriately
contracted over, the subsequent contraction over an $\alpha$ index will yield a term like
$(J_2 \hat{A}_{x,y-1} + J_1 \hat{A}_{x,y}) \hat{B}_{x+1,y}$. For clarity, in the spirit of the
example in Eq.~\eqref{eqn:gMPOex}, a typical contraction over the $\beta_{y-1}$ index
(with $\hat{C}$ = $\hat{0}$) would look like,
\begin{gather}
    \sum_{\beta_{y-1}} \hat{O}_{\beta_{y-1}}[x,y-1] ~ \hat{M}_{\beta_{y-1}}[x,y]  = \nonumber \\
    \hat{I} \cdot
    \left(\begin{array}{cccc}
    \hat{I}_{x,y} & \hat{0} & \hat{0} & \hat{0} \\
    \hat{B}_{x,y} & \hat{0} & \hat{0} & \hat{0} \\
    \hat{0} & \hat{0} & \hat{0} & \hat{0} \\
    \hat{0} & J_1 \hat{A}_{x,y} & \hat{B}_{x,y} & \hat{I}_{x,y}
    \end{array}\right)
    +  \nonumber \\
    \hat{A}_{x,y-1} \cdot
    \left(\begin{array}{cccc}
    \hat{0} & \hat{0} & \hat{0} & \hat{0} \\
    \hat{0} & \hat{0} & \hat{0} & \hat{0} \\
    J_2 \hat{I}_{x,y} & \hat{0} & \hat{0} & \hat{0} \\
    J_1 \hat{B}_{x,y} & J_2 \hat{I}_{x,y} & \hat{0} & \hat{0}
    \end{array}\right) \nonumber \\
    =
    \left(\begin{array}{cccc}
    \hat{I} & \hat{0} & \hat{0} & \hat{0} \\
    \hat{B}_{x,y} & \hat{0} & \hat{0} & \hat{0} \\
    J_2 \hat{A}_{x,y-1} & \hat{0} & \hat{0} & \hat{0} \\
    J_1 \hat{A}_{x,y-1} \hat{B}_{x,y} & J_1 \hat{A}_{x,y} + J_2 \hat{A}_{x,y-1} & \hat{B}_{x,y}
    & \hat{I}
    \end{array}\right).
\end{gather}

The form of this gMPO, which is the simplest case where $\hat{A}$ can interact
with a $\hat{B}$ from a different row \textit{and} column,
is the basis for generating all the more complicated finite-range 2D Hamiltonians with interactions
between more distant neighbors. In essence, the form of $\hat{M}_1$ has to be adapted to the desired
pattern of operators within the gMPO row, and then $\hat{M}_2, \hat{M}_3, ...,$ etc. take the forms which
properly couple the operators from the vertical MPOs (\texttt{intops})
into $\hat{M}_1$.  For a general construction
of this form that includes
all non-symmetric interactions between neighbors up to range $R$, see Appendix A.

The speed and accuracy of the boundary gMPOs using these tensors (with $\hat{C} = \hat{0}$,
$\hat{A} = \hat{B} = \sigma_z$, $J_2 = J_1/2$) is compared to a PEPO-based
implementation and a brute force implementation in Fig.~\ref{fig:no_coeff_acc} and
Tables~\ref{tab:PEPOspeedups}-\ref{tab:bruteSpeedups}. The gMPOs produce accuracies which
are nearly identical to the brute force scheme, but with a computational effort that is
$\sim 60-70\times$ less.
When compared to PEPOs, a speedup of up to $\sim 50\times$ is observed and in most
cases the gMPOs and PEPOs also produce the same level of accuracy. In cases
where they differ, the gMPOs are observed to be more accurate.

\subsection{Long-range Hamiltonians with no coefficients}
\label{sec:LRNC}
We will now consider a Hamiltonian which has local 1-body terms and
non-symmetric pairwise interactions of equal strength
between every site on the lattice. This can be viewed as the 2D version of the Hamiltonian represented
by $\hat{W}_{uniform}$ (see Section~\ref{sec:expMPO}). We have,
\begin{equation}
    \hat{H}_{LRNC} = \sum_i \hat{C}_i + \sum_{i<j} \hat{A}_i \hat{B}_j.
    \label{eqn:ham-2d-lrnc}
\end{equation}
The MPO used in step~\ref{step:botMPO} of the boundary gMPO algorithm
is given by $\hat{W}_{uniform}$. The vertical MPOs
used for the construction of \texttt{intops} are given by,
\begin{gather}
    \hat{O}_{\beta_1}[x,1] =
    \left( \begin{array}{cc}
    \hat{I} & \hat{A}
    \end{array}\right), \nonumber \\
    \hat{O}_{\beta_{y-1} \beta_y}[x, L_y > y > 1] = \left( \begin{array}{cc}
    \hat{I} & \hat{A} \\
    \hat{0} & \hat{I}
    \end{array}\right).
    \label{eqn:intopsLRNC}
\end{gather}
These MPO matrices differ from those in Eq.~\eqref{eqn:intopsTensors1} because they ``remember'' the action
of all the local operators in a given column $x$. In Eq.~\eqref{eqn:intopsTensors1}, the contractions over
$\beta_{y-1}$ that are performed in step~\ref{step:intops} result in operator vectors of the form
$\left( \hat{I}, ~ \hat{A}_{x,y} \right)$. This was sufficient because the previous Hamiltonians under
consideration were local, so the action of the $\hat{A}_{x,y-1}, \hat{A}_{x,y-2}, ...,$ etc. operators
had already been completely accounted for by the time the reference row was shifted up by one. However,
in our current Hamiltonian the interactions are long-ranged, so the action of all the local operators
in a given column must be accounted for in a single \texttt{intops} tensor. This is achieved by
the MPO matrices in Eq.~\eqref{eqn:intopsLRNC}, for which a contraction over
$\beta_1, \beta_2, ..., \beta_{y-1}$
yields operator vectors of the form
$\left( I, ~ \hat{A}_{x,1} + \hat{A}_{x,2} + ... + \hat{A}_{x,y} \right)$.

The corresponding gMPO tensors are given by,
\begin{gather}
    \hat{M}_1[1,y>1] = \left( \begin{array}{cccc}
    \hat{C} & \hat{A} & \hat{B} & \hat{I}
    \end{array}\right), \nonumber \\
    \hat{M}_1[L_x,y>1] = \left( \begin{array}{cccc}
    \hat{I} & \hat{B} & \hat{0} & \hat{C}
    \end{array} \right) ^{T}, \nonumber \\
    \hat{M}_1[L_x > x > 1, y>1] = \left( \begin{array}{cccc}
    \hat{I} & \hat{0} & \hat{0} & \hat{0} \\
    \hat{B} & \hat{I} & \hat{0} & \hat{0} \\
    \hat{0} & \hat{0} & \hat{I} & \hat{0} \\
    \hat{C} & \hat{A} & \hat{B} & \hat{I}
    \end{array} \right), \nonumber \\
    \hat{M}_2[1,y>1] = \left( \begin{array}{cccc}
    \hat{B} & \hat{I} & \hat{0} & \hat{0}
    \end{array}\right), \nonumber \\
    \hat{M}_2[L_x,y>1] = \left( \begin{array}{cccc}
    \hat{0} & \hat{0} & \hat{I} & \hat{B}
    \end{array} \right) ^{T}, \nonumber \\
    \hat{M}_2[L_x > x > 1, y>1] = \left( \begin{array}{cccc}
    \hat{0} & \hat{0} & \hat{0} & \hat{0} \\
    \hat{0} & \hat{0} & \hat{0} & \hat{0} \\
    \hat{I} & \hat{0} & \hat{0} & \hat{0} \\
    \hat{B} & \hat{I} & \hat{0} & \hat{0}
    \end{array} \right).
    \label{eqn:gMPOLRNC}
\end{gather}
Note that this result is nearly identical to the gMPO tensors in the previous section
for diagonal interactions (Eq.~\eqref{eqn:gMPOdiag}).
The only difference is the replacement of two $\hat{0}$s with $\hat{I}$s in $\hat{M}_1$ (and the removal of the
interaction coefficients). The reason for this similarity can be understood in two distinct ways.
Firstly, the addition of these identities can be viewed as an elevation of the symmetric
nearest-neighbor interactions in $\hat{W}_{NN-sym}$ to symmetric interactions of arbitrary range, which captures
all the new terms in $\hat{H}_{LRNC}$.
Secondly, we can see a direct analogy between the relations of the current $\hat{M}_1$ to
$\hat{W}_{uniform-sym}$ (Section~\ref{sec:expMPO})
and the previous $\hat{M}_1$ (Eq.~\eqref{eqn:gMPOdiag}) to $\hat{W}_{NN-sym}$.
In other words, in the previous
section we argued that because $\hat{M}_1$ only differed
from $\hat{W}_{NN-sym}$ by a single element, it
was clear that it would
encode the symmetric nearest-neighbor action of $\hat{B}$ about site $(x,y)$ that was necessary to
generate the diagonal interactions. Now in the current case, we replace the
modified $\hat{W}_{NN-sym}$ with
an identically modified $\hat{W}_{uniform-sym}$ to obtain the symmetric action of
$\hat{B}$ on \textit{all} sites
to the left and right of $(x,y)$. This is precisely the pattern of operators that needs to be encoded
in order to generate all the terms in $\hat{H}_{\mathrm{bot}}$.

The performance of the boundary gMPOs using these tensors (with $\hat{C} = \hat{0}$,
$\hat{A} = \hat{B} = \sigma_z$) is compared to a PEPO-based
implementation and a brute force implementation in Fig.~\ref{fig:no_coeff_acc} and
Tables~\ref{tab:PEPOspeedups}-\ref{tab:bruteSpeedups}.
\REVISION{In this case, due to the long-range
nature of the interactions, the scaling of our brute force evaluation is $O(N^3)$. While
this can be slightly reduced with appropriate caching of contraction intermediates, 
the gMPO- and PEPO-based techniques only scale as $\sim O(N)$ (where $N$ is the total number of sites in
the system).} Thus in addition to the $\sim 60\times$ speedup over the PEPOs, the gMPOs attain large
speedups of $\sim 600\times$ over the brute force algorithm for the $N=64$ cases that we consider. For larger
systems, this speedup will grow rapidly. Given this poor scaling and the fact that the gMPOs
can reproduce the accuracy of the brute
force calculations in all of the most challenging test cases,
it is clear that the brute force technique is not a viable approach to
study systems with non-local interactions.
Of the two viable strategies, gMPOs show very similar accuracy to
PEPOs across most of the test cases, as in the previous sections.

\subsection{Long-range isotropic Hamiltonians with approximate coefficients}
\label{sec:LRAC}

In the previous section, we demonstrated an exact and compact representation of
a long-range interacting 2D Hamiltonian when the interactions coefficients were all the same
(this can also be done with a PEPO~\cite{o2018efficient}). Despite this,
it is a challenging problem
to efficiently~\footnote{Here we define ``efficient'' to mean that the computational cost to
evaluate the expectation value of the Hamiltonian scales linearly with the number of sites in the
system.} represent a 2D Hamiltonian which has long-range
interaction coefficients that depend on the distance between sites, even in an approximate
manner~\cite{frowis2010tensor,o2018efficient,li2019generalization,lin2019low}. Various
solutions to this problem have been proposed
recently~\cite{o2018efficient,li2019generalization,lin2019low},
but they all require the explicit use of PEPOs, making their computational
cost high.

The introduction of the gMPO formalism allows for a new, simpler approach to be derived, which we will
show to be many orders of magnitude more accurate and efficient than the PEPO-based approaches.
We will consider a restricted case of the general long-range interacting Hamiltonian on the 2D lattice,
\begin{equation}
    \hat{H}_{LRAC} = \sum_i \hat{C}_i + \sum_{i<j} V_{ij} \hat{A}_i \hat{B}_j,
    \label{eqn:ham-2d-lrac}
\end{equation}
where $V$ is a translation invariant, decaying function of the Euclidean distance between sites $i$ and $j$ (i.e. it is isotropic).

The crux of the long-range interaction problem on the 2D lattice is that functions of the Euclidean distance
$f(\sqrt{x^2 + y^2})$, which are necessary for physical potentials $V$, are
difficult to represent efficiently within a tensor network structure~\cite{o2018efficient,li2019generalization}.
Although 1D functions of $x$ and $y$
can be independently constructed with ease (see Section~\ref{sec:expMPO}), the known
possibilities for
combining them within a 2D tensor network ansatz yield functions of the Manhattan
distance $f(|x|+|y|)$ or product functions $f(x) g(y)$, but not the desired
radially symmetric ones $f(r) = f(\sqrt{x^2 + y^2})$.
However, the Gaussian function
$e^{-\lambda r^2}$ has the unique property that $f(x)f(y) = e^{-\lambda (x^2 + y^2)} = f(r^2)$.
This connection allows for a radially symmetric Gaussian function in 2D to be created from the product of
two 1D Gaussians $f(x) = e^{-\lambda x^2}$ and $f(y) = e^{-\lambda y^2}$.

This observation can be directly exploited by the gMPO-based algorithm. If the vertical MPOs encode
the interactions $\hat{H}_1 = \sum_{i=1}^{L_y} \sum_{j>i} e^{-\lambda (j-i)^2}
(\hat{A}_i \hat{I}_j + \hat{A}_i \hat{B}_j)$
and the gMPOs encode horizontal interactions of the form
$\hat{H}_2 = \sum_{i=1}^{L_x} \sum_{j > i} e^{-\lambda (j-i)^2} (\hat{I}_i \hat{B}_j +
\hat{B}_i \hat{I}_j + \hat{A}_i \hat{B}_j)$, then they can be combined as a product (as in
Sections~\ref{sec:diag}, \ref{sec:LRNC}) to make complete interactions
of the form $e^{-\lambda (a^2 + b^2)} \hat{A}_{x,y} \hat{B}_{x+a, y\pm b} +
e^{-\lambda a^2} \hat{A}_{x,y} \hat{B}_{x+a, y} + e^{-\lambda b^2} \hat{A}_{x,y} \hat{B}_{x, y+b}$.
The two-dimensional, radially symmetric
Gaussians can then be used as a basis to fit the desired long-range potential,
\begin{equation}
    V(x,y) \approx \sum_{k=1}^{K} c_k e^{-\lambda_k (x^2 + y^2)},
\end{equation}
which is a well-studied problem with highly accurate, compact solutions when $V$ smoothly decays with
distance~\cite{hackbusch2005,beylkin2005,beylkin2010}. The
expectation value of the desired Hamiltonian can then be
evaluated as the sum over the expectation values obtained using
 $K$ different sets of vertical MPOs and
gMPOs (for the $K$ different values of $\lambda$).
Since the only requirement of this technique is the representation of
1D Gaussian functions, this basis can be encoded directly within the MPO and gMPO tensors, which
completely avoids the conceptual and computational
complexity of introducing fictitious superlattices,
as in Refs.~\cite{o2018efficient,li2019generalization}.

\begin{figure}[ht]
    \centering
    \begin{tabular}{c}
    \includegraphics[width=0.45\textwidth]{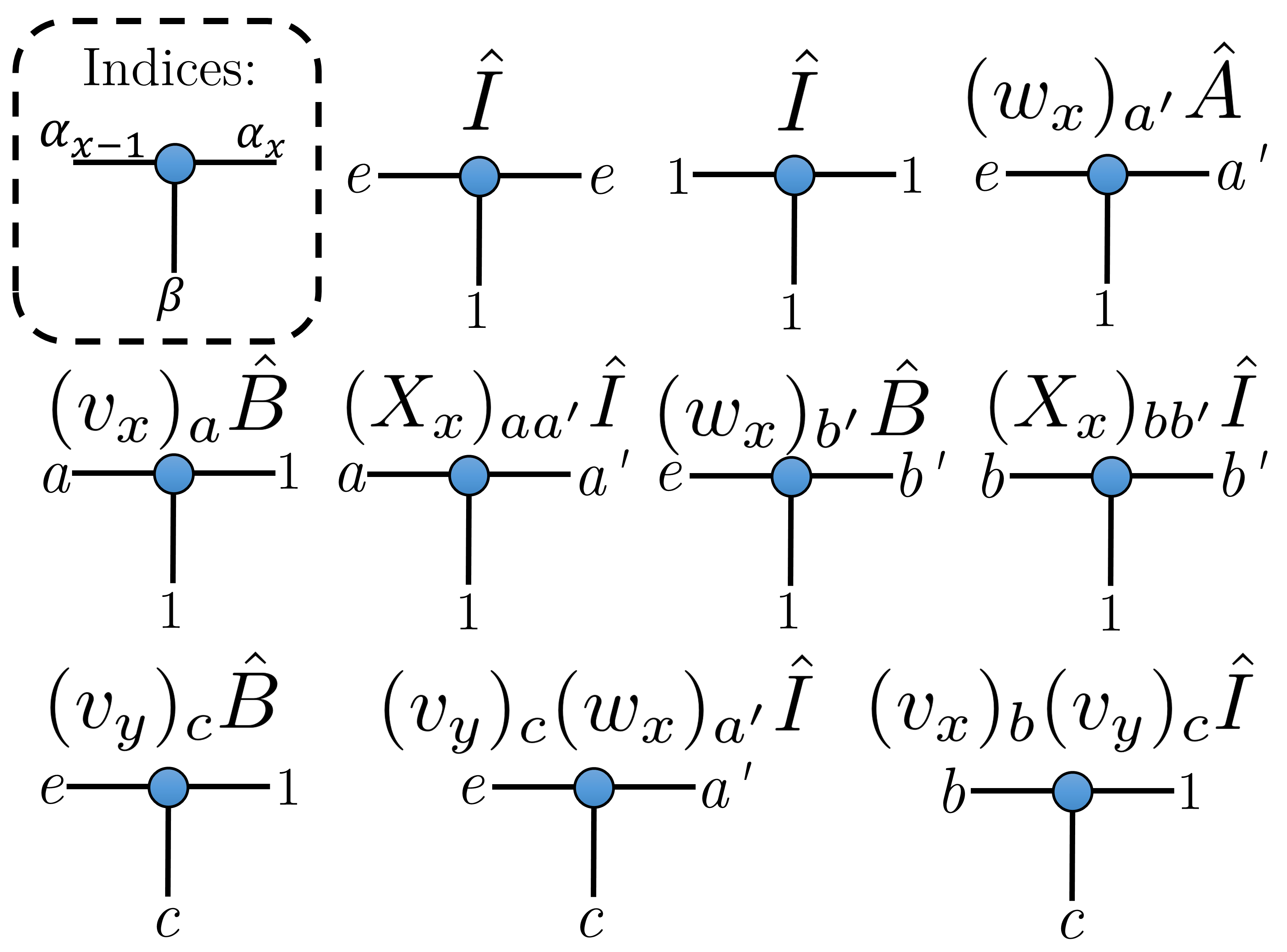} \\
    (a) \\
    \includegraphics[width=0.45\textwidth]{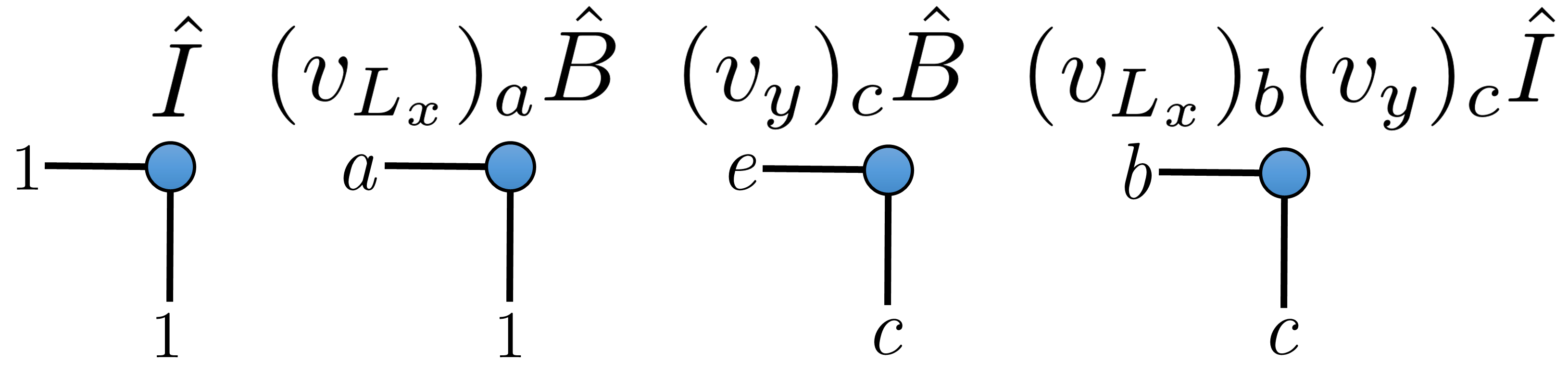} \\
    (b) \\
    \includegraphics[width=0.45\textwidth]{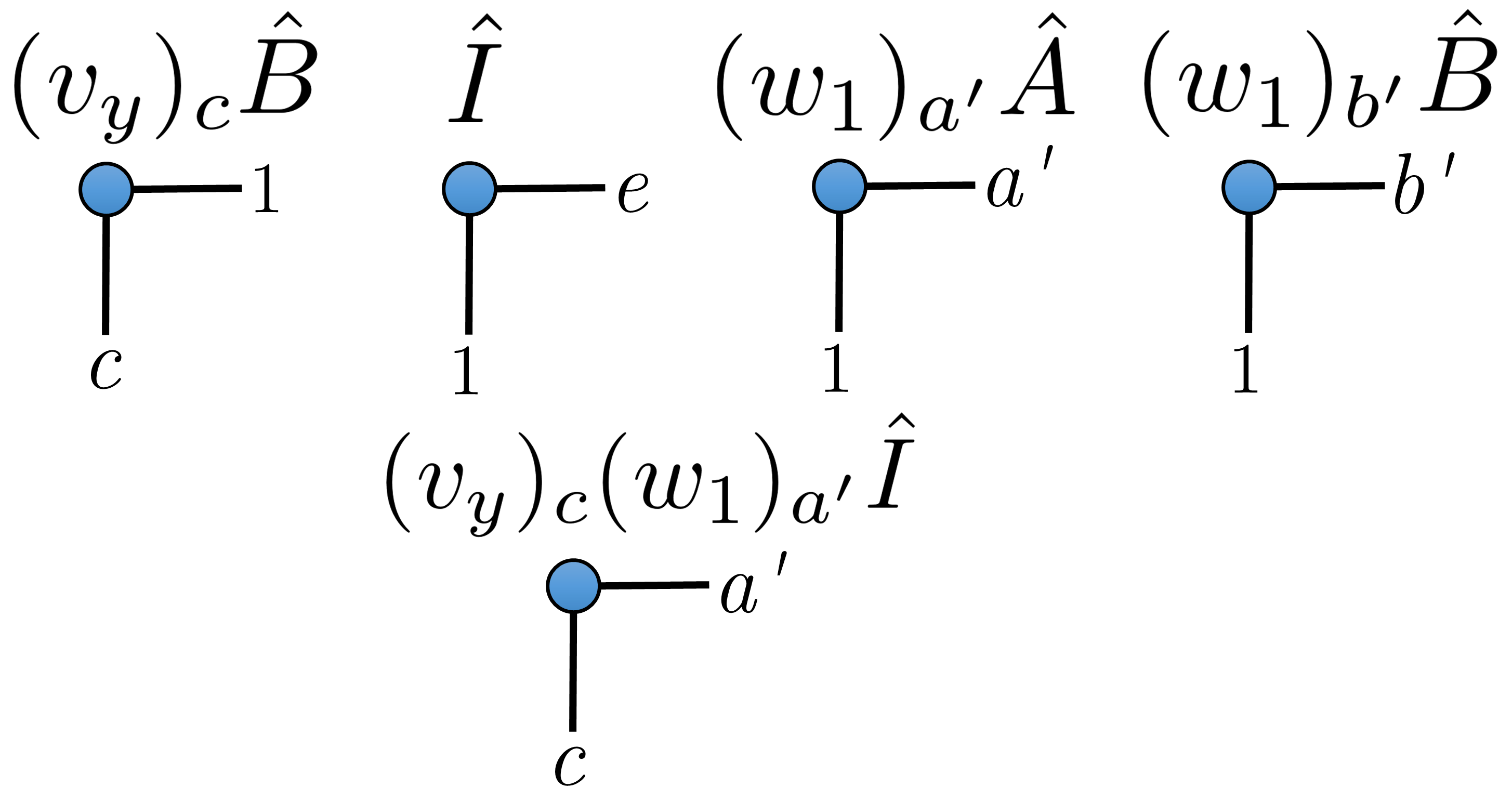} \\
    (c)
    \end{tabular}
    \caption{The gMPO tensors $\hat{M}[x,y>1]$ for long-range Gaussian interactions
    (Eq.~\eqref{eqn:gMPOLRAC}).
    (a) Is for $1<x<L_x$, (b) is for $x=L_x$, and
    (c) is for $x=1$. Since the dimension of each bond $\alpha_{x-1}, \alpha_x$, and
    $\beta$ can vary depending on $x$, $y$ and the
    value of the exponential coefficient $\lambda$ in the Gaussian interaction,
    we will use symbols to label
    specific values of the indices. For a horizontal ($\alpha$) index of bond dimension $2g+2$
    that takes index values
    $\{1, 2, ..., 2g+2 \}$, we label the first value by ``1'',
    the next $g$ values
    by $a$ (if the bond points left) or $a'$ (if the bond points right), the next
    $g$ values by $b$ (if the bond points left) or $b'$ (if the bond points right),
    and the final value by $e$.
    This is the convention that is explained in Eq.~\eqref{eqn:generalMPOsymm} and is also
    used in Eq.~\eqref{eqn:gMPOLRAC}. For a vertical ($\beta$) index of
    bond dimension $g+1$ that takes
    index values $\{1,2,...,g+1 \}$,
    we label the first element by ``1'' and the remaining $g$ elements by $c$. This
    corresponds directly with Eqns.~\eqref{eqn:intopsLRAC} \& \eqref{eqn:gMPOLRAC}
    }
    \label{fig:LRACtensors}
\end{figure}

\begin{figure}[t]
    \centering
    \begin{tabular}{c}
    \includegraphics[width=0.48\textwidth]{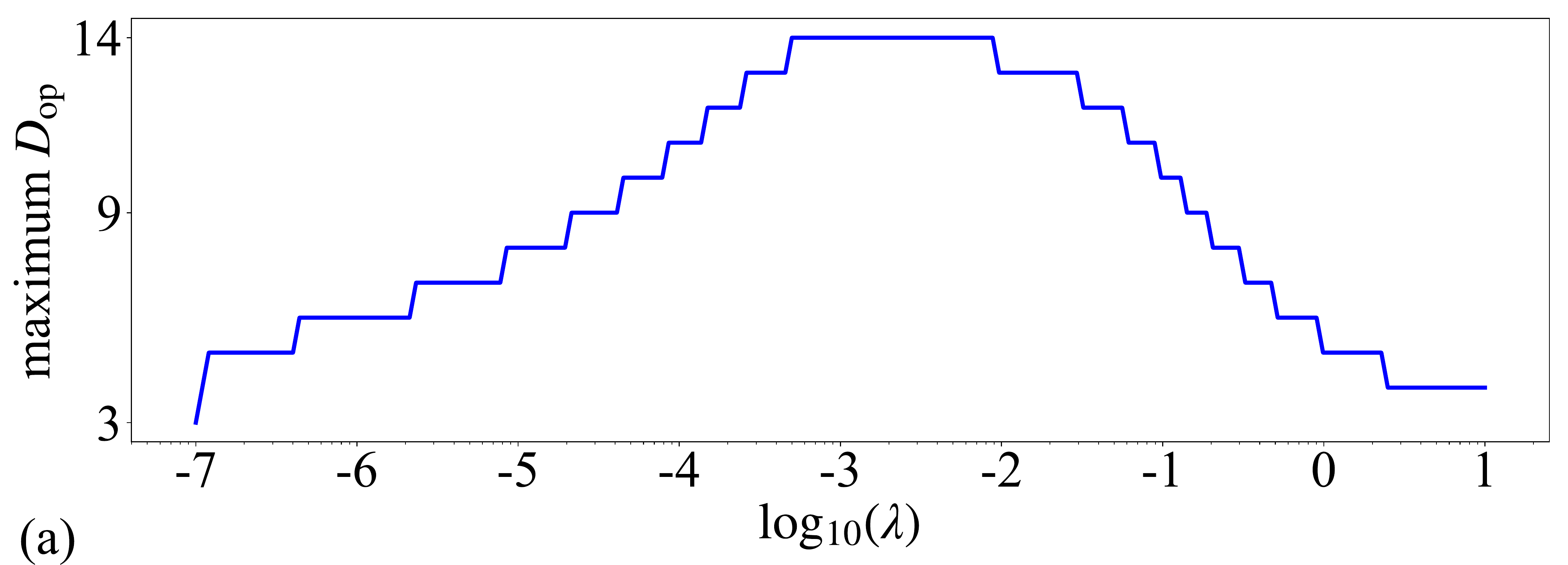} \\
    \includegraphics[width=0.48\textwidth]{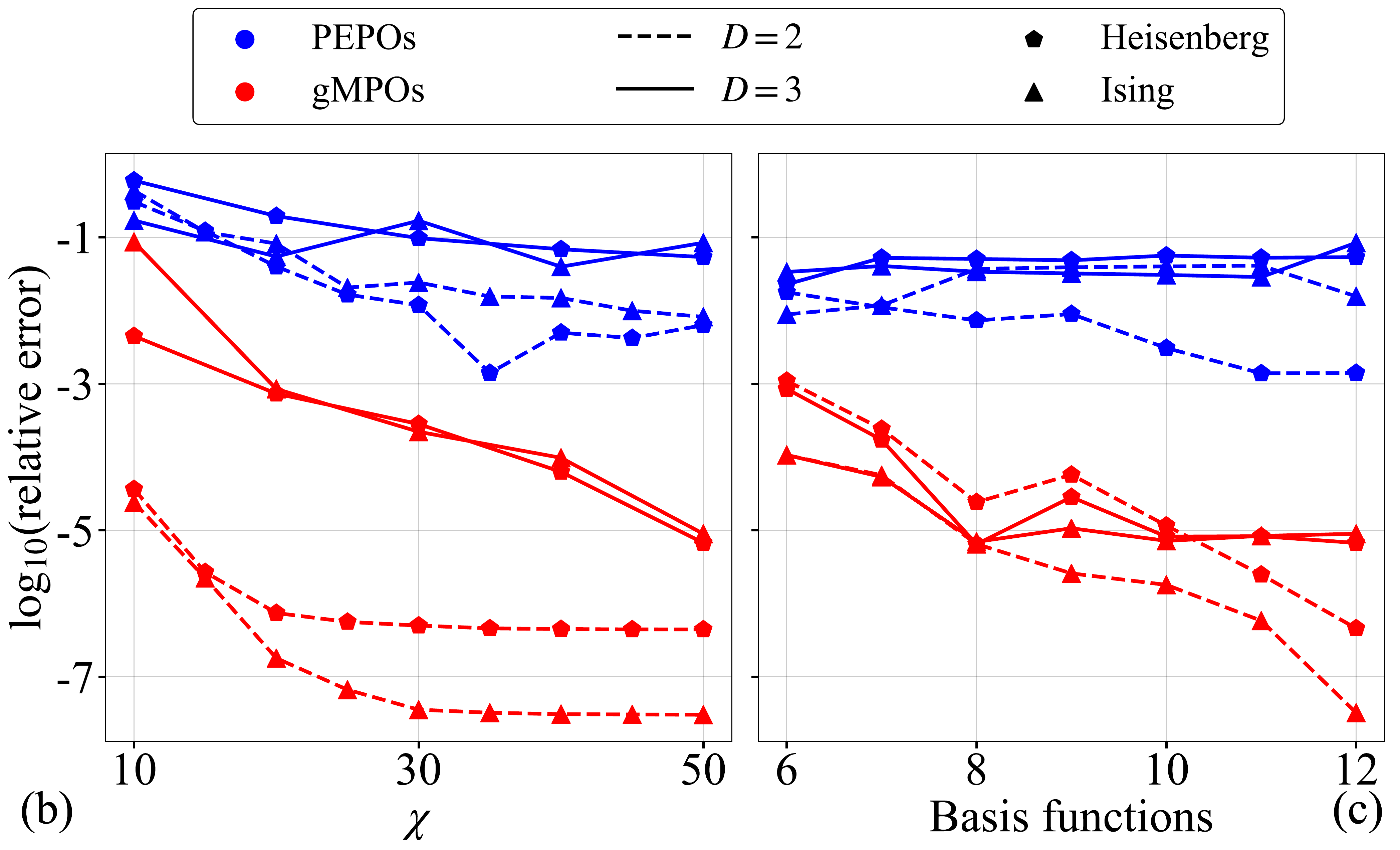}
    \end{tabular}
    \caption{
    (a) The maximum bond dimension of a numerically exact $L=250$ MPO representation of long-range
    pairwise Gaussian interactions for various values of the exponential coefficient $\lambda$. The
    algorithm from~\cite{slicedbasis} was used with a singular value threshold of $10^{-10}$.
    (b)-(c) The relative error in the computed
    expectation value of $\hat{H} = \sum_{i<j} \sigma^z_i \sigma^z_j / |\vecr_i - \vecr_j|$ for
    $8 \times 8$, $D=2$ (dashed) and 3 (solid) ground states
    of the AFM Heisenberg model (pentagons) and FM transverse field Ising model (triangles).
    We compare the Gaussian gMPO technique (red) with the CF-PEPO technique (blue) from
    Ref.~\cite{o2018efficient}. In (b) we use 12 basis functions to fit the Coulomb potential and
    vary the boundary dimension $\chi$ of the contraction algorithm~\cite{lubasch_unifying}. In (c)
    we fix $\chi=35$ for $D=2$, $\chi=50$ for $D=3$ and vary the number of basis functions $K$.
    The convergence of the gMPOs is rapid and strictly governed by $K$ and $\chi$ (for a given
    trial state, either the curve in (b) or in (c) is always decreasing), while the PEPOs
    converge slowly and become saturated by other sources of numerical error.
    }
    \label{fig:LRAC}
\end{figure}

Unfortunately, there is no known exact, compact representation of a 1D MPO with pairwise Gaussian interactions.
However, it can be generated in a nearly numerically exact manner using the method outlined in
Section~\ref{sec:LRMPO} to create $\hat{W}_{gen}$.
Fig.~\ref{fig:LRAC}(a) shows the required bond dimension for the Gaussian
MPO for different values of $\lambda$. The result that $D_{\mathrm{op}} = 14$ in the worst case
for an accuracy of $\sim 10^{-10}$
is a modest bond dimension for an MPO,
which is what makes the current approach of using an exact Gaussian basis
amenable to the gMPO algorithm. Although this same scheme could,
in principle, be implemented using PEPOs on the same lattice,
it would require the use of PEPOs with $D_{\mathrm{op}} = 28$ in the worst case.
In practice, the factor
of $D_{\mathrm{op}}^7$ in the computational cost of PEPO-based contractions makes
a PEPO with a bond dimension of this
size unusable. However, since the use of gMPOs reduces the dependence of the cost on the operator
bond dimension to at most $D_{\mathrm{op}}^3$ (in step~\ref{step:gMPO}), and $D_{\mathrm{op}}^1$ in the most
time intensive step (compression in step~\ref{step:intops}),
using this bond dimension for the vertical MPOs and gMPOs is entirely feasible.

The explicit forms of the tensors in this case can be viewed as a direct
generalization of the tensors from the previous Section (\ref{sec:LRNC}),
Eqs.~\eqref{eqn:intopsLRNC} \& \eqref{eqn:gMPOLRNC}.
This follows from the discussion in Section~\ref{sec:LRMPO} regarding
$\hat{W}_{gen}$ and $\hat{W}_{gen-sym}$ as
direct generalizations of $\hat{W}_{uniform}$ and $\hat{W}_{uniform-sym}$,
respectively. Since the tensors in Section~\ref{sec:LRNC} are
derived from $\hat{W}_{uniform(-sym)}$ and in the current case we want to use tensors based on
the $\hat{W}_{gen(-sym)}$ representation of a Gaussian MPO, the tensors in~\eqref{eqn:intopsLRNC}
and \eqref{eqn:gMPOLRNC} generalize to the current case in an
analogous way to the $\hat{W}_{uniform(-sym)} \to \hat{W}_{gen(-sym)}$
generalization of Section~\ref{sec:LRMPO}.

Specifically, the MPO for step~\ref{step:botMPO} is the $\hat{W}_{gen}$ representation of Gaussian
interactions with exponential coefficient $\lambda_k$, which is determined from the algorithm in
Ref.~\cite{slicedbasis}. From this MPO, the data for each $\vec{v}_i$, $\vec{w}_i$, and $X_i$ can
be extracted (according to Eq.~\eqref{eqn:generalMPO}). These can then be used to construct the
other tensors for pairwise interactions mediated by a 2D Gaussian potential.
The vertical MPO tensors are given by,
\begin{gather}
    \hat{O}_{\beta_1}[x,1] =
    \left( \begin{array}{cc}
    \hat{I} & (w_1)_c \hat{A}
    \end{array}\right), \nonumber \\
    \hat{O}_{\beta_{y-1} \beta_y}[x, L_y > y > 1] = \left( \begin{array}{cc}
    \hat{I} & (w_y)_c \hat{A} \\
    \hat{0} & (X_y)_{c' c} \hat{I}
    \end{array}\right),
    \label{eqn:intopsLRAC}
\end{gather}
where $c$ and $c'$ index through the vector $\vec{w}_y$ and matrix $X_y$,
like in Eq.~\eqref{eqn:generalMPO}. The gMPO tensors are,
\begin{gather}
    \hat{M}_1[1,y>1] = \left( \begin{array}{cccc}
    \hat{C}, & (w_1)_{a'} \hat{A}, & (w_1)_{b'} \hat{B}, & \hat{I}
    \end{array}\right), \nonumber \\
    \hat{M}_1[L_x,y>1] = \left( \begin{array}{cccc}
    \hat{I}, & (v_{L_x})_{a} \hat{B}, &  \hat{0}, & \hat{C}
    \end{array} \right) ^{T}, \nonumber \\
    \hat{M}_1[L_x > x > 1, y>1] = \left( \begin{array}{cccc}
    \hat{I} & \hat{0} & \hat{0} & \hat{0} \\
    (v_x)_a \hat{B} & (X_x)_{a a'} \hat{I} & \hat{0} & \hat{0} \\
    \hat{0} & \hat{0} & (X_x)_{b b'} \hat{I} & \hat{0} \\
    \hat{C} & (w_x)_{a'} \hat{A} & (w_x)_{b'} \hat{B} & \hat{I}
    \end{array} \right), \nonumber \\
    \hat{M}_c[1,y>1] = \left( \begin{array}{cccc}
    (v_y)_c \hat{B}, & (v_y)_c \cdot (w_1)_{a'} \hat{I}, & \hat{0}, & \hat{0}
    \end{array}\right), \nonumber \\
    \hat{M}_c[L_x,y>1] = \left( \begin{array}{cccc}
    \hat{0}, & \hat{0}, & (v_y)_c \cdot (v_{L_x})_{b} \hat{I}, & (v_y)_c \hat{B}
    \end{array} \right) ^{T}, \nonumber \\
    \hat{M}_c[L_x > x > 1, y>1] = \nonumber \\
    \left( \begin{array}{cccc}
    \hat{0} & \hat{0} & \hat{0} & \hat{0} \\
    \hat{0} & \hat{0} & \hat{0} & \hat{0} \\
    (v_y)_c \cdot (v_x)_b \hat{I} & \hat{0} & \hat{0} & \hat{0} \\
    (v_y)_c \hat{B} & (v_y)_c \cdot (w_x)_{a'} \hat{I} & \hat{0} & \hat{0}
    \end{array} \right), \nonumber \\
    c \in \{2,3,..., \texttt{len}( \vec{w}_{y-1} ) + 1 \}.
    \label{eqn:gMPOLRAC}
\end{gather}
Here $c$ is used consistently between Eqns.~\eqref{eqn:intopsLRAC} \& \eqref{eqn:gMPOLRAC}
to index the vertical MPO bond $\beta$.
In an identical manner to Eq.~\eqref{eqn:generalMPOsymm}, $a$, $a'$, $b$, $b'$ are used to
index the coefficient vectors $\vec{w}_x$, $\vec{v}_x$ and the coefficient matrix $X_x$.
Note that within a given $\hat{M}_c$ matrix, the value of $c$ is fixed while the values of
$a$, $a'$, $b$, $b'$ range appropritely over the dimensions of the matrix. This means that
an expression such as $(v_y)_c \cdot (v_x)_b$ in Eq.~\eqref{eqn:gMPOLRAC}
is a scalar multiplying a vector.

For these expressions to always make sense, we require
$L_x \geq L_y$ so that the bottom MPO is long enough to extract all the necessary coefficient
vectors and matrices for the vertical direction. The crucial component of this representation
is how $\vec{w}$ and $\vec{v}$ appear in $\hat{M}_c$. For the $\hat{I}$
operator in the bottom row of the matrix, which
couples the action of $\hat{A}$ (from below) to the action of
$\hat{B}$ (to the right) in the gMPO row, the ``completion'' interaction
coefficients $\vec{v}_y$ are encoded along the $\beta$ index while the ``beginning'' interaction
coefficients $\vec{w}_x$ are encoded along the $\alpha_x$ index. Similarly for the $\hat{I}$
operator in the first column of the matrix,
which couples the action of $\hat{A}$ (from below) to the action
of $\hat{B}$ (to the left) in the gMPO row, the ``completion'' interaction
coefficients $\vec{v}_y$ are encoded along the $\beta$ index while the ``completion'' interaction
coefficients $\vec{v}_x$ are encoded along the $\alpha_{x-1}$ index. This formulation
allows for vertical interactions of the form
$\sum_{i=1}^y e^{-\lambda (y-i)^2} \hat{A}_i \hat{I}_y$ to be ``completed'' and thus
scalar multiplied by ``completed'' horizontal interactions of the form
$\sum_{i=x+1}^{L} e^{-\lambda (i-x)^2} \hat{I}_x \hat{B}_i$ and
$\sum_{i=1}^x e^{-\lambda (x-i)^2} \hat{B}_i \hat{I}_x$,
yielding the desired 2D Gaussian potential. The other entries of the tensors can
be understood by their analogous form to the previous section and their direct
correspondence with $\hat{W}_{gen-sym}$ (Eq.~\eqref{eqn:generalMPOsymm}).

Due to the inherent challenge of explicitly writing and interpreting
the algebraic expressions for $\hat{M}$ when the dimension of the $\beta$ index
is greater than 2, it can be more intuitive to understand the form of these gMPO tensors from
a graphical presentation, which is given in Fig.~\ref{fig:LRACtensors}. Additionally,
a straightforward example implementation of these tensors can be
found online~\cite{linkToGithub}.

The performance of this scheme relative to the PEPO-based scheme from
Ref.~\cite{o2018efficient}
for evaluating the expectation value of $\hat{H} = \sum_{i<j} (\sigma^z_i \sigma^z_j)/|\vecr_i-\vecr_j|$
is shown in Fig.~\ref{fig:LRAC} and Table~\ref{tab:PEPOspeedups}.
One notable difference between this case and the previous sections is
that there is no longer a generic, guaranteed speedup of the gMPOs over PEPOs
because the two methods work differently. The PEPOs encode long-range coefficients by
introducing a large auxiliary lattice, while the gMPOs do so by using an increased bond
dimension. Since these things affect the computational scaling in different ways and their
precise costs depend on specific numerical thresholds,
one method is not strictly faster than the other.


However, in practice we observe that the gMPOs are many
orders of magnitude more computationally efficient than the PEPOs. The simplest way to see this
is to first note that for given values of $K$, $\chi$, and $D>2$, the CF-PEPO and gMPO schemes
require similar levels of computational effort (Table~\ref{tab:PEPOspeedups}). Yet with these same
parameters, the gMPOs are approximately 4 orders of magnitude more accurate than the CF-PEPOs
(Fig.~\ref{fig:LRAC}). This can be extended to recognize that in order to obtain
a given level of accuracy,
the gMPOs will be many orders of magnitude faster than the PEPOs, or more generally that the
gMPOs can obtain a more accurate answer than the PEPOs in less time.

Additionally, the convergence towards high accuracy is faster and more straightforward
when using gMPOs than when using PEPOs. In the case of the gMPOs, the accuracy is systematically
governed by $\chi$ and $K$ (see Fig.~\ref{fig:LRAC}(b)-(c)). This becomes clear by
observing that, for a given trial state, its curve in either Fig.~\ref{fig:LRAC}(b) or (c)
is always decreasing. On the other hand, the convergence of the PEPO curves stall.
The medium- and high-accuracy regimes are not bounded by errors due to
the basis size or $\chi$, but instead by larger numerical errors stemming from additional
complicated parameters involved with making the basis radially
symmetric~\cite{o2018efficient,li2019generalization}. In fact,
this is the inherent reason for the major accuracy difference. The gMPO Gaussian basis
is radially symmetric up to $\sim 10^{-10}$ (the singular value threshold used in the
approximation algorithm),
whereas the PEPO bases are only radially symmetric up to significant numerical
errors~\cite{o2018efficient,li2019generalization}.

As a final point, we note that a slightly faster implementation of this long-range gMPO
scheme is possible.
Since the bond dimensions $D_{\mathrm{op}}$ reported in Fig.~\ref{fig:LRAC}(a) are only for
$\hat{W}_{gen}$, the horizontal
bond dimension of the gMPO tensors in Eq.~\eqref{eqn:gMPOLRAC} is almost twice as large. A factor
of $\sim 2$ speedup can be gained in step~\ref{step:gMPO} of the boundary gMPO algorithm
if non-symmetric gMPO tensors of horizontal dimension $D_{\mathrm{op}}$ are used instead,
so that the interactions
$\hat{A}_{x,y} \hat{B}_{x,y+b} + \hat{A}_{x,y} \hat{B}_{x+a,y} +
\hat{A}_{x,y} \hat{B}_{x+a,y+b}$ are encoded in one gMPO and
the interactions $\hat{A}_{x,y} \hat{B}_{x-a,y+b}$ in another. The cost of this bond dimension
reduction is an increase in the number of gMPOs that need to be independently evaluated from
$K$ to $2K$, but this still leaves a factor of 2 for the speedup because
the cost of step~\ref{step:gMPO} depends quadratically on the horizontal bond dimension
of the gMPOs.


\section{Conclusions}
\label{sec:conclude}
In this work we have presented an algorithm which can evaluate the expectation value of general
2D Hamiltonians without using a PEPO. To accomplish this, we introduced the formalism of a
gMPO and showed how it can be used in combination with MPOs to efficiently compute the
energy of a PEPO on-the-fly. In addition to the conceptual simplification of rewriting PEPOs
in terms of the more familiar MPOs, we also showed that computing the energy using this strategy is
1-2 orders of magnitude faster while being equally as accurate as explicitly using a PEPO.
The structure of the algorithm also allows for a new technique to be used for constructing and
evaluating 2D Hamiltonians with physical long-range interaction potentials,
which we demonstrated to be
multiple orders of magnitude more accurate and efficient than existing strategies.
We expect that this work will lower the computational and conceptual barriers to using
tensor network operators in future PEPS calculations. We hope that this opens the door to the
study of new, more complicated Hamiltonians in the tensor network community.

Finally, although this work focused on the specific case of finite systems, the fundamental requirement for
the formulation of the algorithm to apply is that the contraction method starts from the boundary.
Since much is known about infinite
MPOs~\cite{mcculloch2008infinite,zauner2018vumps,parker2019local}
and many prominent contraction methods for infinite PEPS~\cite{jordan2008classical} also begin from
the boundary~\cite{orus2009simulation,corboz2016variational,vanderstraeten2016gradient,fishman2018},
we expect that the concepts presented in this work can be generalized to the infinite case.

\section*{Acknowledgements}
Primary support for this work was from AFOSR MURI Grant FA9550-18-1-0095. M.J.O. acknowledges
financial support from a US National Science Foundation Graduate Research Fellowship via Grant
DEG-1745301. G.K.C. acknowledges support from the Simons Foundation. The authors thank Henry
Schurkus and Zhendong Li for helpful feedback on the manuscript.

\section*{Appendix A: MPO and gMPO for general finite-range Hamiltonian}
In Section~\ref{sec:diag}, we reported the exact construction of the vertical MPO matrices
and the gMPO tensors for a Hamiltonian that had non-symmetric ``linear'' interactions up to distance
$R=1$, and non-symmetric diagonal interactions up to distance $R=\sqrt{2}$. Following the concepts
in that example, and the general ideas behind MPO construction, in this Appendix
we will give the exact construction for a Hamiltonian with non-symmetric linear interactions up to a general
distance $R$, and non-symmetric diagonal interactions up to $\sqrt{2}R$. The interactions coefficients
will be denoted $j_{x,y}$, where $x$ is the horizontal distance between the local operators and $y$ is the
vertical distance.

The vertical MPO matrices are size $(R+2) \times (R+2)$ and they are given by,
\begin{gather}
    \hat{O}_{\beta_1}[x,1] =
    \left( \begin{array}{ccc}
    \hat{I} &  \hat{A} & \begin{bmatrix} \hat{0} \end{bmatrix}_{R-1}
    \end{array}\right), \nonumber \\
    \hat{O}_{\beta_{y-1} \beta_y}[x, y > 1] =
    \left( \begin{array}{c|c|c}
    \hat{I} & \hat{A} &\hat{0} \\
    \hline
    \hat{0} & \hat{0} & \mathbb{1}_{(R-1) \times (R-1)} \hat{I} \\
    \hline
    \hat{0} & \hat{0} & \hat{0}
    \end{array} \right).
\end{gather}

The gMPO tensors are size $(2R+2) \times (2R+2) \times (R+1)$, where the third dimension
is the size of the $\beta$ index. They are given by,
\begin{gather}
    \hat{M}_1[1,y>1] = \nonumber \\
    \left( \begin{array}{ccccccc}
    \hat{C}, & j_{1,0} \hat{A}, & \cdots, & j_{R,0} \hat{A}, & \hat{B}, &
    \begin{bmatrix} \hat{0} \end{bmatrix}_{R-1} & \hat{I}
    \end{array}\right), \nonumber \\
    \hat{M}_1[1,y>1] =
    \left( \begin{array}{cccc}
    \hat{I}, & \hat{B}, &
    \begin{bmatrix} \hat{0} \end{bmatrix}_{2R-1} & \hat{C}
    \end{array}\right)^T, \nonumber \\
    \hat{M}_1[L_x > x > 1, y>1] = \nonumber \\
    \left( \begin{array}{c|c|c|c|c|c}
    \hat{I} & \hat{0} & \hat{0} & \hat{0} & \hat{0} & \hat{0} \\
    \hline
    \hat{B} & \hat{0} & \hat{0} & \hat{0} & \hat{0} & \hat{0} \\
    \hline
    \hat{0} & \mathbb{1}_{(R-1) \times(R-1)} \hat{I} & \hat{0} & \hat{0} & \hat{0} & \hat{0} \\
    \hline
    \hat{0} & \hat{0} & \hat{0} & \hat{0} & \mathbb{1}_{(R-1) \times(R-1)} \hat{I} &  \hat{0} \\
    \hline
    \hat{0} & \hat{0} & \hat{0} & \hat{0} & \hat{0} & \hat{0} \\
    \hline
    \hat{C} &
    \begin{matrix}
    j_{1,0} \hat{A}, & \cdots, & j_{R-1,0} \hat{A}
    \end{matrix} & j_{R,0} \hat{A} & \hat{B} & \hat{0} & \hat{I}
    \end{array} \right), \nonumber \\
    \hat{M}_k[1,y>1] = \nonumber \\
    \left( \begin{array}{ccccc}
    j_{0,k-1} \hat{B}, & j_{1,k-1} \hat{I}, & \cdots, & j_{R,k-1} \hat{I}, &
    \begin{bmatrix} \hat{0} \end{bmatrix}_{R+1}
    \end{array}\right), \nonumber \\
    \hat{M}_k[L_x,y>1] = \nonumber \\
    \left( \begin{array}{ccccc}
    \begin{bmatrix} \hat{0} \end{bmatrix}_{R+1} & j_{1,k-1} \hat{I}, & \cdots, & j_{R,k-1} \hat{I}, & j_{0,k-1} \hat{B}
    \end{array} \right) ^{T}, \nonumber \\
    \hat{M}_k[L_x > x > 1, y>1] = \nonumber \\
    \left( \begin{array}{c|c|c}
    \begin{bmatrix} \hat{0} \end{bmatrix}_{R+1} & \hat{0} & \hat{0} \\
    \hline
    \begin{matrix}
    j_{1,k-1} \hat{I} \\
    j_{2,k-1} \hat{I} \\
    \vdots \\
    j_{R,k-1} \hat{I}
    \end{matrix} & \hat{0} & \hat{0} \\
    \hline
    j_{0,k-1} \hat{B} & \begin{matrix}
    j_{1,k-1} \hat{I}, & j_{2,k-1} \hat{I}, & \cdots, & j_{R,k-1} \hat{I}
    \end{matrix}
    & \begin{bmatrix} \hat{0} \end{bmatrix}_{R+1}
    \end{array} \right), \nonumber \\
    k \in \{2,3,...,R+1 \}.
\end{gather}
In these expressions, $\mathbb{1}_{N \times N}$ is an $N\times N$ identity matrix.
Additionally, when something is enclosed in square brackets and labelled with a
subscript $n$, it means ``repeat this $n$ times''. Based on the dimensions of the other
blocks, it should be clear which axis it should be expanded along.
This is only used in places where it is not otherwise obvious to expand the blocks
to match the dimensions of adjacent blocks.



%

\end{document}